\documentclass[
    a4paper,
    notitlepage,
]{revtex4-1}
\usepackage{amsmath,amssymb}
\usepackage{lipsum,soul}
\usepackage{bm}
\usepackage{graphicx}
\usepackage{graphics}
\usepackage{amsmath}
\usepackage{array}
\usepackage[caption=false]{subfig}
\usepackage{xcolor}
\usepackage{subfig}
\usepackage{hyperref}
\usepackage[capitalise]{cleveref}
\usepackage{textcomp} 

\usepackage{epstopdf}
\usepackage{cases}
\usepackage{amssymb}
\usepackage{multirow}
\usepackage{siunitx}
\usepackage{cases}
\usepackage{tabularx}
\usepackage{nomencl}
\usepackage[utf8]{inputenc}
\usepackage{tikz}
\usepackage[T1]{fontenc}
\usepackage{comment}
\usepackage{lineno,hyperref}
\usepackage{booktabs}

\begin{document}

\title{Heat vortexes of ballistic, diffusive and hydrodynamic phonon transport in two-dimensional materials}
\author{Chuang Zhang}
\email{zhangcmzt@hust.edu.cn}
\affiliation{%
 State Key Laboratory of Coal Combustion, School of Energy and Power Engineering, Huazhong University of Science and Technology, Wuhan 430074, China}%
\author{Songze Chen}
\email{jacksongze@hust.edu.cn}
\affiliation{%
 State Key Laboratory of Coal Combustion, School of Energy and Power Engineering, Huazhong University of Science and Technology, Wuhan 430074, China}%
\author{Zhaoli Guo}%
\email{Corresponding author: zlguo@hust.edu.cn}
\affiliation{%
 State Key Laboratory of Coal Combustion, School of Energy and Power Engineering, Huazhong University of Science and Technology, Wuhan 430074, China}%

\date{\today}

\begin{abstract}

In this work, the heat vortexes in two-dimensional porous or ribbon structures are investigated based on the phonon Boltzmann transport equation (BTE) under the Callaway model.
First, the separate thermal effects of normal (N) scattering and resistive (R) scattering are investigated with frequency-independent assumptions.
And then the heat vortexes in graphene are studied as a specific example.
It is found that the heat vortexes can appear in both ballistic (rare R/N scattering) and hydrodynamic (N scattering dominates) regimes but disappear in the diffusive (R scattering dominates) regime.
As long as there is not sufficient R scattering, the heat vortexes can appear in present simulations.

\end{abstract}

\maketitle

\section{INTRODUCTION}

Heat conduction in bulk materials at room temperature is usually described by the Fourier's law of thermal conduction, which implies that the heat carriers (phonons) undergo a diffusive process~\cite{kaviany_2008,ChenG05Oxford,liu_anomalous_2014}.
However, as the system size or dimension decreases, the phonons no longer transport diffusively~\cite{RevModPhysJoseph89,nanoscaleenergy2020,RevModPhysLibaowen,RevModPhys.90.041002,chang_breakdown_2008,PhysRevB.100.085203,PhysRevLett.107.095901,ZHANG2020,PhysRevB.101.075303,de_tomas_kinetic_2014,WangMr15application} due to the existence of the ballistic~\cite{volz1999,MajumdarA93Film} or hydrodynamic~\cite{beck1974,PhysRevLett.28.1461,PhysRevLett_secondNaF,PhysRevLett_ssNaf,machida2018,PhysRev_GK,lee_hydrodynamic_2015,cepellotti_phonon_2015,huberman_observation_2019} phonon transport.

The ballistic phonon transport happens as the system characteristic length is comparable to or smaller than the phonon mean free path~\cite{MajumdarA93Film,ChenG05Oxford,kaviany_2008}.
In the ballistic regime, the phonon transport is blocked by boundary scattering so that the thermal conductivity decreases significantly as the system size decreases~\cite{volz1999,MajumdarA93Film,li_thermal_2003,ju1999phonon,xu_length-dependent_2014}.

Different from the ballistic~\cite{volz1999,MajumdarA93Film} and diffusive phonon transport, the hydrodynamic phonon transport~\cite{Gurzhi_1968,PhysRevB.10.3546,ENZ1968114,beck1974,PhysRevLett.28.1461,PhysRevLett_secondNaF,PhysRevLett_ssNaf,machida2018,PhysRev_GK,lee_hydrodynamic_2015,cepellotti_phonon_2015,huberman_observation_2019} happens as the momentum-conservation phonon-phonon scattering dominates the heat conduction.
Usually, the phonon-phonon scattering is separated into two parts: the momentum destroying umklapp (U) scattering and the momentum-conservation normal (N) scattering~\cite{PhysRevB.102.094311,PhysRevB_SECOND_SOUND,lee_hydrodynamic_2015,cepellotti_phonon_2015,PhysRevB.99.085202}.
In bulk materials at room temperature, the U scattering happens frequently so that the phonon transport recovers the Fourier's law of thermal conduction.
However, in extremely low temperature or low-dimensional materials~\cite{Gurzhi_1968,ENZ1968114,RevModPhys.90.041002,PhysRevLett_Strontium_Titanate,
huberman_observation_2019,PhysRevLett.28.1461,PhysRevLett_secondNaF,PhysRevLett_ssNaf,machida2018}, the N scattering becomes much stronger than the U scattering~\cite{beck1974,PhysRevB.10.3546,PhysRev_GK,RevModPhys.90.041002,lee_hydrodynamic_2015,cepellotti_phonon_2015,PhysRevB.99.085202} and the phonon may transport like the fluid dynamics~\cite{prandtl_fluid,liu_new_2016vortex,GuoZl13LB}.

In the past decades, three phonon hydrodynamic phenomena have been predicted both theoretically~\cite{PhysRevB_SECOND_SOUND,lee_hydrodynamic_2015,cepellotti_phonon_2015,PhysRevB.99.085202,PhysRevB.99.144303,nanoletterchengang_2018} and experimentally~\cite{Dreyer1993,machida2020,PhysRevLett_Strontium_Titanate,
huberman_observation_2019,PhysRevLett.16.789,PhysRevLett.28.1461,PhysRevLett_secondNaF,PhysRevLett_ssNaf,machida2018}, such as the second sound~\cite{RevModPhysJoseph89,Dreyer1993,huberman_observation_2019,PhysRevLett_secondNaF,PhysRevB_SECOND_SOUND,PhysRevLett.28.1461,nie2020thermal}, phonon poiseuille flow~\cite{machida2018} and Knudsen minimum~\cite{PhysRevB.99.085202,nanoletterchengang_2018}.
The second sound is that the transient heat propagates like sound wave because the N scattering recovers the temperature wave equation~\cite{nie2020thermal,lee_hydrodynamic_2015,PhysRev_GK,PhysRev.148.766,huberman_observation_2019}.
Different from the second sound, the phonon poiseuille flow is a steady problem.
Namely, under a small temperature gradient, the heat flux flows in an infinite long and finite width ribbon like the fluid flows in a long tube with parabolic distributions~\cite{machida2018,PhysRevB.99.085202}.
In addition, as the ribbon width decreases, there is a minimum value of the non-dimensional thermal conductivity, which is the phonon Knudsen minimum~\cite{PhysRevB.99.085202,nanoletterchengang_2018}.

So except above three phenomena, can we find another novel phonon hydrodynamic phenomena?
It is well known that the transport of the heat carriers including phonon, electron and gas, can be described by the Boltzmann transport equation with similar mathematical form, although their physical meanings are different~\cite{ChenG05Oxford,kaviany_2008}.
In addition, as normal scattering dominates the heat conduction, the corresponding macroscopic Guyer-Krumhansl (G-K)  equation~\cite{PhysRev_GK,PhysRev.148.766,PhysRevB.10.3546,shang_heat_2020,PhysRevX.10.011019,WangMr15application} is similar to the Navier-Stokes equation in fluid dynamics~\cite{prandtl_fluid,GuoZl13LB}.
Hence, will phonon transport like the vortexes in fluid dynamics~\cite{prandtl_fluid,GuoZl13LB,liu_new_2016vortex} or viscous electron flow in hall bar geometries~\cite{levitov_electron_2016,bandurin2016b,chandra2019a,Danz_2020}?

Recently, the heat vortexes phenomena (\cref{vortexcurl}(a)) in the phonon hydrodynamic regime have been predicted in graphene ribbon~\cite{shang_heat_2020}.
However, a systematic comparison of the heat vortexes in the phonon hydrodynamic, ballistic and diffusive regimes is lacking.
Besides, the frequency dependent N/R scattering is not accounted in previous study~\cite{shang_heat_2020}.

In this work, motivated by the viscous fluid dynamics~\cite{prandtl_fluid,GuoZl13LB,liu_new_2016vortex} or viscous electron flow~\cite{levitov_electron_2016,bandurin2016b,chandra2019a,Danz_2020}, the heat vortexes in two different structures (porous~(\cref{porous}) and ribbon~(\cref{ribbon})) are studied in different regimes based on gray model~\cite{ChenG05Oxford}.
Note that the gray model is not limited by a specific material so that all variables are set to be dimensionless.
It is found that the heat vortexes can appear in both the ballistic and hydrodynamic regimes while disappear in the diffusive regime.
Namely, as long as the R scattering is not sufficient, the heat vortex can appear.
To make results more convincing, graphene~\cite{RevModPhys.90.041002,xu_length-dependent_2014,shang_heat_2020,lee_hydrodynamic_2015,cepellotti_phonon_2015} is taken as a specific example and the heat vortexes in graphene are investigated with different system sizes or temperatures accounting for the phonon dispersion and polarization.

The paper is organized as follows.
In Sec.~\ref{sec:BTE}, the phonon BTE is introduced briefly.
Then a theoretical analysis of the heat vortexes in different regimes is made in Sec.~\ref{sec:possibility}.
In Sec.~\ref{sec:results}, the heat vortexes in two-dimensional porous (\cref{porous}) or ribbon (\cref{ribbon}) structures are investigated.
Finally, a conclusion is made in Sec.~\ref{sec:conclusion}.

\section{Phonon BTE}
\label{sec:BTE}

In order to predict the thermal transport in two-dimensional materials correctly~\cite{RevModPhys.90.041002,xu_length-dependent_2014,lee_hydrodynamic_2015,cepellotti_phonon_2015}, the phonon Boltzmann transport equation (BTE) under the Callaway’s dual relaxation model~\cite{PhysRev_callaway,wangmr17callaway,PhysRevB.99.085202,luo2019}  is used, i.e.,
\begin{equation}
\frac{\partial f}{\partial t }+ \bm{v} \cdot \nabla f= \frac{f^{eq}_{R}-f}{\tau_{R}} + \frac{f^{eq}_{N}-f}{\tau_{N}},
\label{eq:BTE}
\end{equation}
where $f=f(\bm{x}, \omega,p, \bm{K},t)$ is the phonon distribution function. $\bm{x}$ is the spatial position.
$\omega$ is the phonon angular frequency, $p$ is the phonon polarization, $\bm{K}$ is the 2D wave vector and assumed to be isotropic, i.e., $\bm{K}= |\bm{K}|\bm{s}$, $\bm{s}$ is the unit directional vector in 2D coordinate system, $t$ is the time.
$\bm{v} =\nabla _{\bm{K}} \omega$ is the group velocity calculated by the phonon dispersion~\cite{PhysRevLett.95.096105,PhysRevB.79.155413,ChenG05Oxford},
$\tau_{R}$ is the effective relaxation time of the resistive (R) scattering, which is a combination of all momentum destroying phonon scattering except the boundary scattering based on the Mathiessen's rule~\cite{ChenG05Oxford,kaviany_2008}, such as the umklapp (U) scattering, impurity scattering and isotope scattering~\cite{PhysRev_callaway,wangmr17callaway}.
$f^{eq}_{R}$ is the associated equilibrium state of the R scattering and satisfies Bose-Einstein distribution~\cite{ChenG05Oxford,kaviany_2008}, i.e.,
\begin{align}
f^{eq}_{R}(T)  = \frac{1}{ \exp \left( \frac{\hbar \omega }{k_B T} \right) -1 },  \label{eq:feqresistive}
\end{align}
where $k_B$ is the Boltzmann constant, $\hbar$ is the Planck constant reduced by $2\pi$ and $T$ is the temperature.
Different from the R scattering, the normal (N) scattering conserves momentum.
Its displaced equilibrium distribution function and the effective relaxation time are $f^{eq}_{N}$ and $\tau_{N}$, respectively, where
\begin{align}
f^{eq}_{N}(T,\bm{u})= \frac{1}{ \exp \left( \frac{\hbar \omega - \hbar \bm{K} \cdot \bm{u} }{k_B T} \right) -1 }, \label{eq:feqnormal}
\end{align}
where $\bm{u}$ is the drift velocity.
In addition, the phonon wave nature~\cite{PhysRevB.67.195311,ChenG05Oxford,kaviany_2008} is ignored in present study.

Usually, the phonon BTE (Eq.~\eqref{eq:BTE}) is rewritten into a deviational energy form as follows
\begin{align}
\frac{\partial e}{\partial t }+ \bm{v} \cdot \nabla e= \frac{e^{eq}_{R}-e}{\tau_{R}} + \frac{e^{eq}_{N}-e}{\tau_{N}},
\label{eq:energyBTE}
\end{align}
where the associated deviational distribution functions of energy density are~\cite{luo2019,wangmr17callaway}
\begin{align}
e &=\frac{ \hbar \omega  D (f-f_R^{eq} (T_0) ) }{2 \pi}, \\
e_R^{eq} &=\frac{ \hbar \omega  D (f_R^{eq}-f_R^{eq} (T_0) ) }{2 \pi}, \label{eq:eeqresistive} \\
e_N^{eq} &=\frac{ \hbar \omega  D (f_N^{eq}-f_R^{eq} (T_0) ) }{2 \pi}, \label{eq:eeqnormal}
\end{align}
where $D=|\bm{K}|/\left( 2 \pi |\bm{v}| \right)$ is the phonon density of states~\cite{luo2019,wangmr17callaway}.

Assuming a small temperature difference $\Delta T$ and a small drift velocity, i.e., $\Delta T/T_0 \ll 1$, $\bm{K} \cdot \bm{u} \ll \omega$, where $T_0$ is the reference temperature, then the equilibrium distribution functions (Eqs.~\eqref{eq:eeqresistive} and~\eqref{eq:eeqnormal}) can be linearized, i.e.,
\begin{align}
e^{eq}_{R}(T)  &\approx  C \left( T-T_0 \right) / 2\pi, \label{eq:feqR} \\
e^{eq}_{N}(T,\bm{u}) &\approx  C \left( T-T_0 \right) / 2\pi +CT \frac{\bm{K} \cdot \bm{u} }{ 2 \pi \omega},\label{eq:feqN}
\end{align}
where $C=C(\omega,p,T_0)$ is the mode specific heat at $T_0$, i.e.,
\begin{equation}
C(\omega,p,T_0)=2 \pi  \left.\ \frac{  \partial{ e^{eq}_{R}}}{ \partial{T}} \right|_{T=T_0} .
\label{eq:specificheat}
\end{equation}

The local temperature $T$ and heat flux $\bm{q}$ are calculated by taking the moments of the distribution function, i.e.,
\begin{align}
T  &=T_0+ \frac{ \sum_{p} \int \int e d\Omega d\omega } { \sum_{p} \int C d\omega  },   \label{eq:T}  \\
\bm{q} &=  \sum_{p} \int \int \bm{v} e d\Omega d\omega ,
\label{eq:heatflux}
\end{align}
where $p$ is the phonon polarization, $d\Omega$ and $d\omega$ are the integral over the whole 2D solid angle space and frequency space.
In addition, based on the conservation properties of the R (only energy conservation) and N (energy and momentum conservation) scattering, we have
\begin{align}
0  &=  \sum_{p} \int \int  \frac{e^{eq}_{R}(T_R) -e}{\tau_{R}}   d\Omega d\omega , \label{eq:Rconsertvation} \\
0  &=  \sum_{p} \int \int    \frac{e^{eq}_{N} (T_N)-e}{\tau_{N}}   d\Omega d\omega , \label{eq:Nconsertvation}  \\
0  &=  \sum_{p} \int \int  \frac{\bm{K} }{\omega } \frac{e^{eq}_{N} (T_N,\bm{u})-e}{\tau_{N}}   d\Omega d\omega , \label{eq:NMconsertvation}
\end{align}
where two local pseudo-temperatures $T_R$ and $T_N$ are introduced to ensure the conservation principle of the R and N scattering~\cite{WangMr15application,wangmr17callaway}.

By choosing specific heat $C$, reference temperature $T_0$, system characteristic length $L$ and group velocity $\bm{v}$ as the reference variables, the stationary phonon BTE (Eq.~\eqref{eq:energyBTE}) can be normalized as
\begin{align}
\frac{\bm{v} }{ |\bm{v}|} \cdot \nabla_{\bm{x^*}} e^* = \beta_R (e^{eq,*}_{R}-e^*) + \beta_N (e^{eq,*}_{N}-e^*), \label{eq:dimensionlessBTE}
\end{align}
where
\begin{align}
e^* &=\frac{e}{ C T_0 }, &\quad x^* &=\frac{x}{L},&\quad e^{eq,*}_R &=\frac{e^{eq}_R }{ C T_0 } , \notag \\
e^{eq,*}_N &=\frac{e^{eq}_N }{ C T_0 },&\quad \beta_N &=\frac{L}{|\bm{v}| \tau_N},& \quad \beta_R &=\frac{L}{|\bm{v}| \tau_R}.
\end{align}
It can be found that the heat conduction is totally determined by $\beta_N$ and $\beta_R$, which represent the ratio between the system size $L$ to the phonon mean free path of N and R scattering, i.e., $|\bm{v}|\tau_N$ and $|\bm{v}|\tau_R$, respectively.

\section{Theoretical analysis of heat vortexes in the ballistic, diffusive and hydrodynamic regimes}
\label{sec:possibility}

\begin{figure}
 \centering
 \includegraphics[scale=0.30,viewport=0 100 900 500,clip=true]{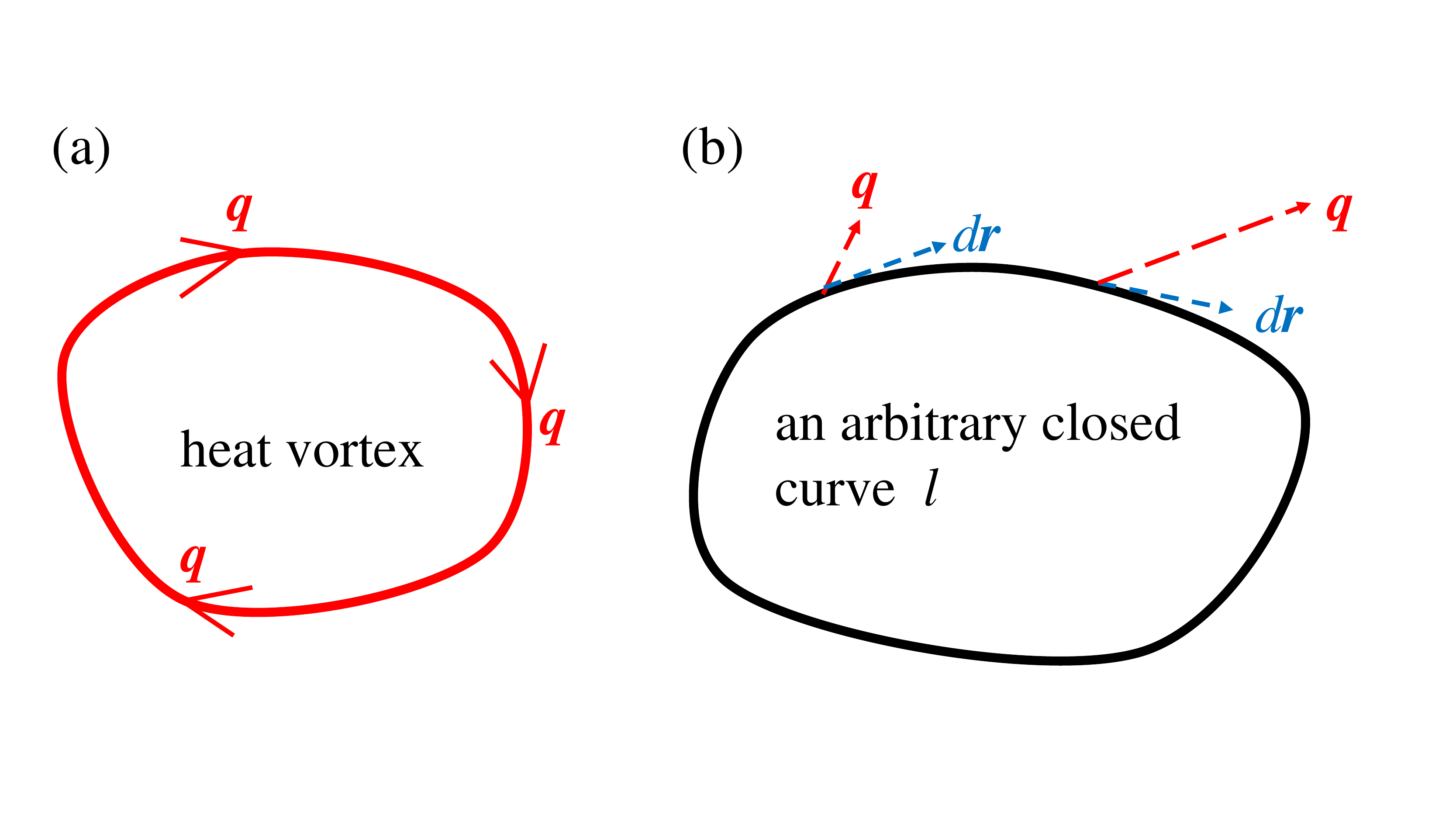}
 \caption{(a) Schematic of the heat vortex~\cite{shang_heat_2020}, where $\bm{q}$ is the heat flux. (b) Given an arbitrary closed curve $l$, the curl of heat flux is $
\int_l \bm{q} \cdot d \bm{r} $~\cite{liu_new_2016vortex,prandtl_fluid}, where $d \bm{r}$ is the unit tangent vector of the closed curve $l$ in a clockwise direction.}
 \label{vortexcurl}
\end{figure}
The heat vortexes (\cref{vortexcurl}(a)) in the diffusive ($\beta_R \rightarrow \infty $, $\beta_N=0$), ballistic ($\beta_R=0$, $\beta_N=0$) and hydrodynamic ($\beta_R=0$, $\beta_N \rightarrow \infty$) limits are analyzed, respectively.
So that the separate thermal effects of R scattering, boundary scattering and N scattering can be found clearly.

For simplicity, the phonon gray model and Debye approximation~\cite{ChenG05Oxford,kaviany_2008} are used, i.e., $\omega= |\bm{v}| |\bm{K}|$.
Taking zero- and first- order moments of phonon BTE (Eq.~\eqref{eq:energyBTE}) leads to~\cite{WangMr15application,PhysRev_GK,PhysRev.148.766,shang_heat_2020}
\begin{align}
\frac{\partial E }{\partial t }+ \nabla \cdot   \bm{q}  &=0, \label{eq:btezero}  \\
\frac{\partial \bm{q}  }{\partial t } +  \nabla   \cdot   \bm{Q}  &= -\frac{\bm{q} }{\tau_R} + 0, \label{eq:rbtefirst}
\end{align}
where
\begin{align}
E & = \sum_{p} \int \int  e  d\Omega d\omega,  \\
\bm{Q} &=  \sum_{p} \int \int  \bm{v} \bm{v} e  d\Omega d\omega,
\end{align}
are the local deviational energy and the macroscopic flux of heat flux, respectively.

In the diffusive limit ($\beta_R \rightarrow \infty$, $\beta_N=0$), the R scattering dominates the phonon transport and causes heat dissipations based on Eq.~\eqref{eq:rbtefirst}.
At steady state, the distribution function can be approximated as
\begin{align}
e \approx e_{R}^{eq}- \tau_R \bm{v} \cdot \nabla e_{R}^{eq} .
\end{align}
So that the Fourier's law of heat conduction can be derived~\cite{GuoZl16DUGKS,ChenG05Oxford,MajumdarA93Film}, i.e.,
\begin{align}
\bm{q} = - \kappa_d \nabla T,
\label{eq:fourier2D}
\end{align}
where $\kappa_d=\frac{1}{2} C | \bm{v} |^2  \tau_R $ is the thermal conductivity in the diffusive limit.

{\color{black}{Note that the vortex definition and identification have
been a longstanding issue with many disputes~\cite{liu_new_2016vortex}, so that here the vorticity is introduced and discussed simply, which is defined as the curl of heat flux in present work~\cite{prandtl_fluid}.
Considering an arbitrary closed curve $l$ inside the thermal system (boundaries are not included), according to Eq.~\eqref{eq:fourier2D}, the curl of heat flux is (\cref{vortexcurl}(b))
\begin{align}
\int_l \bm{q} \cdot  d \bm{r} = -\kappa_d \int_l \frac{dT}{d \bm{r}} \cdot  d \bm{r} =0,   \label{eq:circulationzero}
\end{align}
where $d \bm{r}$ is the unit tangent vector of the closed curve $l$ in a clockwise direction.
This indicates that the curl of heat flux for an arbitrary closed curve inside the thermal system is always zero.
So that it can be concluded that there are no heat vortexes inside the system in the diffusive regime.}}

However, different from diffusive phonon transport, there is rare phonon-phonon scattering in the ballistic limit ($\beta_R=0$, $\beta_N=0$), so that Eqs.~\eqref{eq:energyBTE} and~\eqref{eq:rbtefirst} become,
\begin{align}
\frac{\partial  e   }{\partial t } + \bm{v} \cdot \nabla e  &=0 , \label{eq:bteballistic} \\
\frac{\partial \bm{q} }{\partial t } +  \nabla   \cdot   \bm{Q} &=  0. \label{eq:conserved}
\end{align}
Equation~\eqref{eq:conserved} indicates that there are no heat (or momentum) dissipations.
Besides, Eq.~\eqref{eq:conserved} is also satisfied in the phonon hydrodynamic limit ($\beta_N \rightarrow \infty ,~\beta_R=0$)~\cite{huberman_observation_2019,li2018a,cepellotti_phonon_2015,lee_hydrodynamic_2015}.
Because the N scattering conserves momentum and causes no thermal resistance~\cite{shang_heat_2020,PhysRev.148.766,PhysRev_GK}.
In other words, in both hydrodynamic and ballistic limits, the heat flux (or momentum) is conserved in the interior domain~\cite{Nanalytical,cepellotti_phonon_2015,lee_hydrodynamic_2015}, which is different from that in the diffusive limit.

Although Eq.~\eqref{eq:conserved} is satisfied in the ballistic and phonon hydrodynamic limits, there is something different.
In the ballistic regime~\cite{PhysRevB.99.085202}, the momentum conservation is satisfied due to rare phonon-phonon scattering.
The phonon mean free path is much larger than the local characteristic length.
At steady state, Eq.~\eqref{eq:bteballistic} becomes
\begin{align}
\bm{v} \cdot \nabla  e =0,
\end{align}
which indicates that for a given group velocity, the phonon distribution function does not vary with the spatial position until scattering with the boundaries.
Namely, the heat flux distributions inside the domain, which are obtained by taking the first-order moment of distribution function (Eq.~\eqref{eq:heatflux}), depend on the boundary conditions.
Therefore, whether the heat vortexes appear in the ballistic regime depends on the geometry settings or boundary conditions.

However, in the phonon hydrodynamic regime, the phonon mean free path is much smaller than the local characteristic length due to the frequent N scattering~\cite{Gurzhi_1968,beck1974,PhysRevLett.28.1461,PhysRevLett_secondNaF,PhysRevLett_ssNaf,PhysRev_GK,lee_hydrodynamic_2015,cepellotti_phonon_2015,huberman_observation_2019}.
At steady state, the phonon distribution function can be approximated as~\cite{PhysRev_GK,PhysRev.148.766,WangMr15application,shang_heat_2020}
\begin{align}
e \approx e_{N}^{eq}- \tau_N \bm{v} \cdot \nabla e_{N}^{eq}.
\label{eq:firstCEhydrodynamic}
\end{align}
It can be found that the phonon distribution function varies with spatial position unless $\tau_N =0$ according to Eq.~\eqref{eq:firstCEhydrodynamic}, which is different from that in the ballistic regime.

In addition, the macroscopic Guyer-Krumhansl (G-K) equation of hydrodynamic phonon transport in three-dimensional bulk materials has been derived based on the phonon BTE as $\beta_N \gg 1.0 \gg  \beta_R$ (Eq.~\eqref{eq:dimensionlessBTE})~\cite{PhysRev_GK,PhysRev.148.766,PhysRevB.10.3546,PhysRevX.10.011019,WangMr15application}.
A recent study has also extended it from three-dimensional into two-dimensional materials, i.e.,
(Eq.(33) in Ref~\cite{shang_heat_2020})
\begin{align}
\frac{\partial \bm{q} }{\partial t} + \frac{1}{2} C | \bm{v} |^2  \nabla  T + \frac{\bm{q}}{\tau_R} = \frac{1}{4}  | \bm{v} |^2  \tau_N \left( \nabla ^2 \bm{q}  \right) . \label{eq:GKshang}
\end{align}
If the R scattering is totally ignored, at steady state, Eq.~\eqref{eq:GKshang} becomes
\begin{align}
C   \nabla  T  = \frac{1}{2}  \tau_N \left( \nabla ^2 \bm{q}   \right). \label{eq:steadyGKshang}
\end{align}
It can be found that the right hand side of Eq.~\eqref{eq:steadyGKshang} is similar to the viscous term in Navier-Stokes equation~\cite{prandtl_fluid,GuoZl13LB}, so that it is possible to predict the heat vortexes in the phonon hydrodynamic regimes.
Besides, based on Eq.~\eqref{eq:steadyGKshang}, whether the heat vortexes appear in the phonon hydrodynamic regimes depends on the boundary conditions, too.

In a word, the heat vortexes disappear in the diffusive regime.
But they may appear in both ballistic and hydrodynamic regimes, which depends on the boundary conditions or geometry settings.

\section{Numerical results and discussions}
\label{sec:results}

In order to validate above theoretical analysis, the heat vortexes in two-dimensional porous and ribbon structures are introduced and investigated numerically in different regimes, respectively.
The implicit discrete ordinate method~\cite{ZHANG20191366,wangmr17callaway} is used to solve the phonon BTE, which can refer to Ref~\cite{wangmr17callaway}.
Detailed numerical solutions of the phonon BTE and associated boundary conditions~\cite{luo2019,ZHANG20191366} can be seen in Appendix~\ref{sec:callaway}.

\subsection{Heat vortex in two-dimensional porous materials}

\subsubsection{Problem description}

\begin{figure}
 \centering
 \includegraphics[scale=0.30,viewport=100 0 800 520,clip=true]{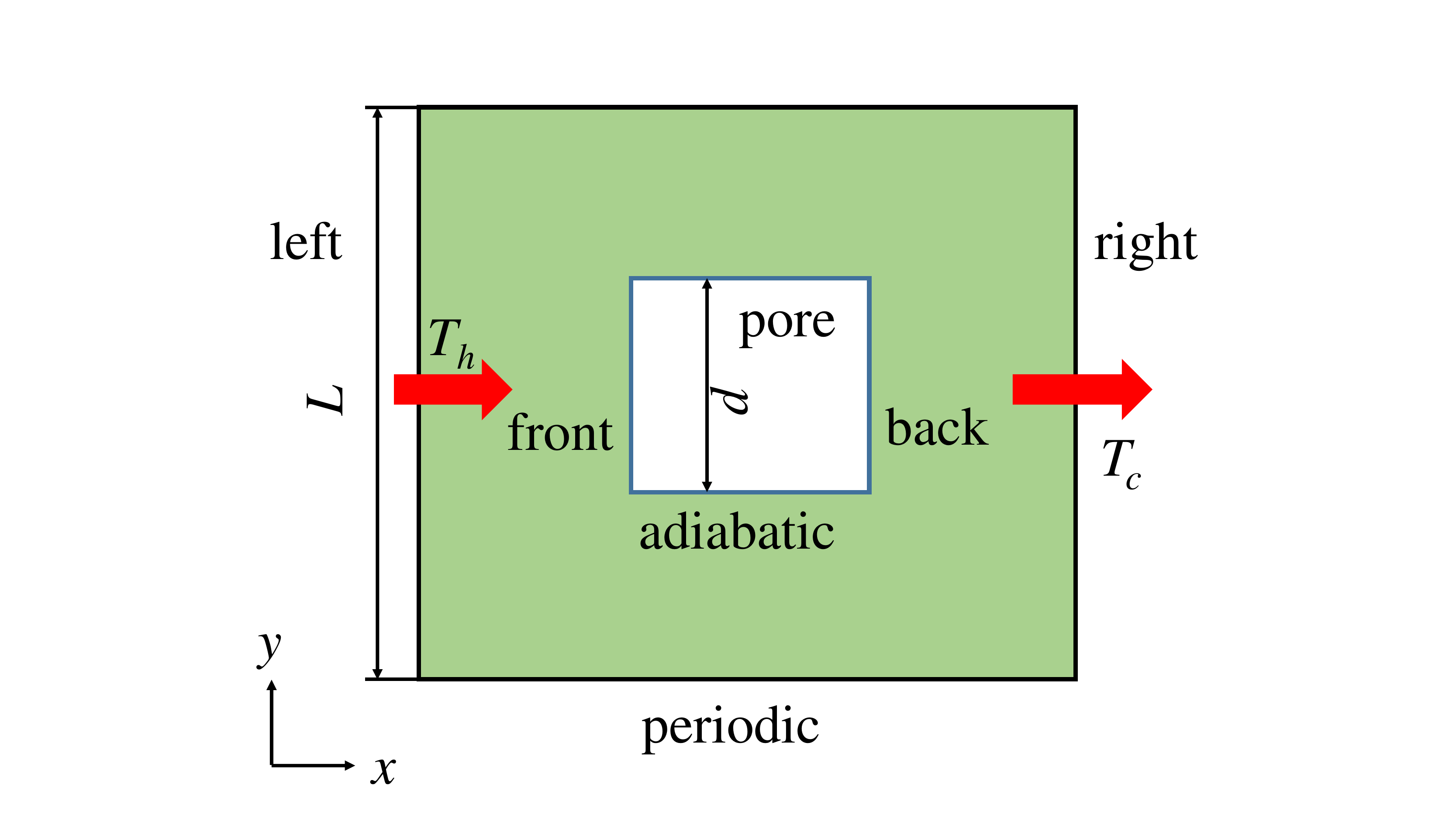}
 \caption{Schematic of a unit 2D square cell with a square pore in the center. The side lengths of the unit cell and square pore are $L$ and $d=L/2$, respectively. The boundaries of the unit cell and the square pore are periodic and adiabatic, respectively. A small temperature gradient is added across the $x$ direction so that the heat flows from the left to the right around the square pore, where $T_h=T_0 +\Delta T/2$, $T_c= T_0 -\Delta T/2$, $\Delta T/T_0=0.01$.}
 \label{porous}
\end{figure}
The heat conduction in periodic two-dimensional porous materials with aligned square pores is studied first.
Figure~\ref{porous} shows one of the unit 2D square cells with side length $L$.
In its center, there is a square pore with side length $d=L/2$.
The temperatures of the left and right sides of the cell are $T_h =T_0 +\Delta T/2$ and $T_c=T_0 - \Delta T/2$, respectively, where $T_0 = \left(T_h +T_c  \right)/2$ and $\Delta T= T_h - T_c$ are the average temperature and temperature difference in the domain.
In what follows, we set $\Delta T/T_0 =0.01$, which leads to a small temperature gradient along the $x$ direction.
The initial temperature in the interior domain is $T_0$.
The boundaries of the cell and the square pore are periodic (Eq.~\eqref{eq:BC2}) and adiabatic (Eq.~\eqref{eq:BC1}), respectively.
When phonons encounter the pore boundaries, they suffer from diffusely reflecting scattering due to the roughness of the pore surface (Eq.~\eqref{eq:BC1}).
So the heat flows from the left to the right around the square pore, which is like the gas flow around a cylinder in fluid dynamics~\cite{GuoZl13LB}.

\subsubsection{Gray model}

The phonon gray model and Debye approximation~\cite{ChenG05Oxford,kaviany_2008} are used. It is not limited by a specific material so that all physical variables are dimensionless.
Based on Eq.~\eqref{eq:dimensionlessBTE} and dimensional analysis~\cite{barenblatt1987dimensional}, as long as $\beta_N$ and $\beta_R$ are fixed, the results predicted by the phonon BTE will be fixed.
For simplicity, in the following simulations, only $\tau_R^{-1}$ or $\tau_N^{-1}$ is changed and all other variables are fixed, i.e., $L=2d=1$, $C=1,~|\bm{v}|=1,~T_0=1$.

\begin{figure}
 \centering
 \includegraphics[scale=1.1,viewport=50 150 500 300,clip=true]{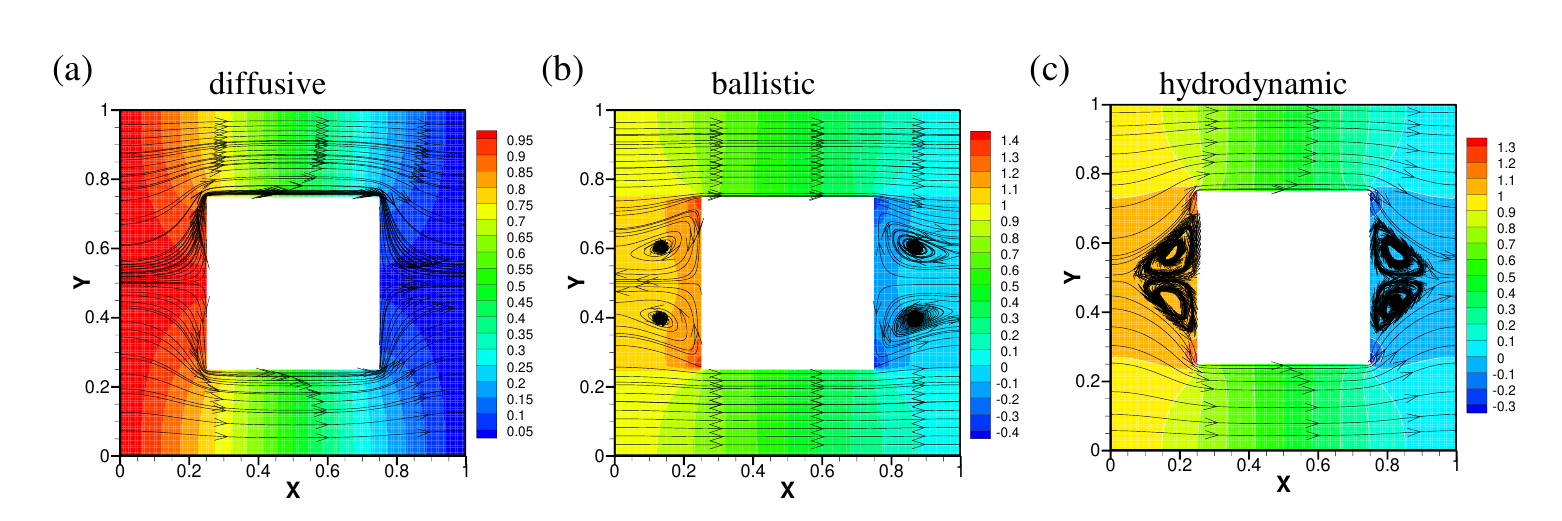}
 \caption{Normalized temperature contour $(T-T_c)/\Delta T$ and heat flux streamline in different regimes in porous structures (\cref{porous}) based on gray model. It is not limited by a specific material so that all physical variables are set to be dimensionless, where $L=2d=1,~C=1,~|\bm{v}|=1,~T_0=1$. $X$ and $Y$ are the coordinates. (a) Diffusive ($\tau_R=0.01,~\tau_N^{-1}=0$), (b) ballistic ($\tau_R^{-1}=0,~\tau_N^{-1}=0$) and (c) hydrodynamic ($\tau_R^{-1}=0,~\tau_N=0.01$).}
 \label{porousBDH}
\end{figure}

Numerical results are shown in~\cref{porousBDH}.
It can be observed that heat vortexes appear in both ballistic and hydrodynamic regimes but disappear in the diffusive regime.
The results are in consistent with our theoretical analysis as mentioned in last section.
Besides, the vortex sizes of the ballistic phonon transport are larger than those in the hydrodynamic regime.

\subsubsection{Porous graphene}

\begin{figure}
     \centering
     \includegraphics[scale=1.1,viewport=50 150 500 300,clip=true]{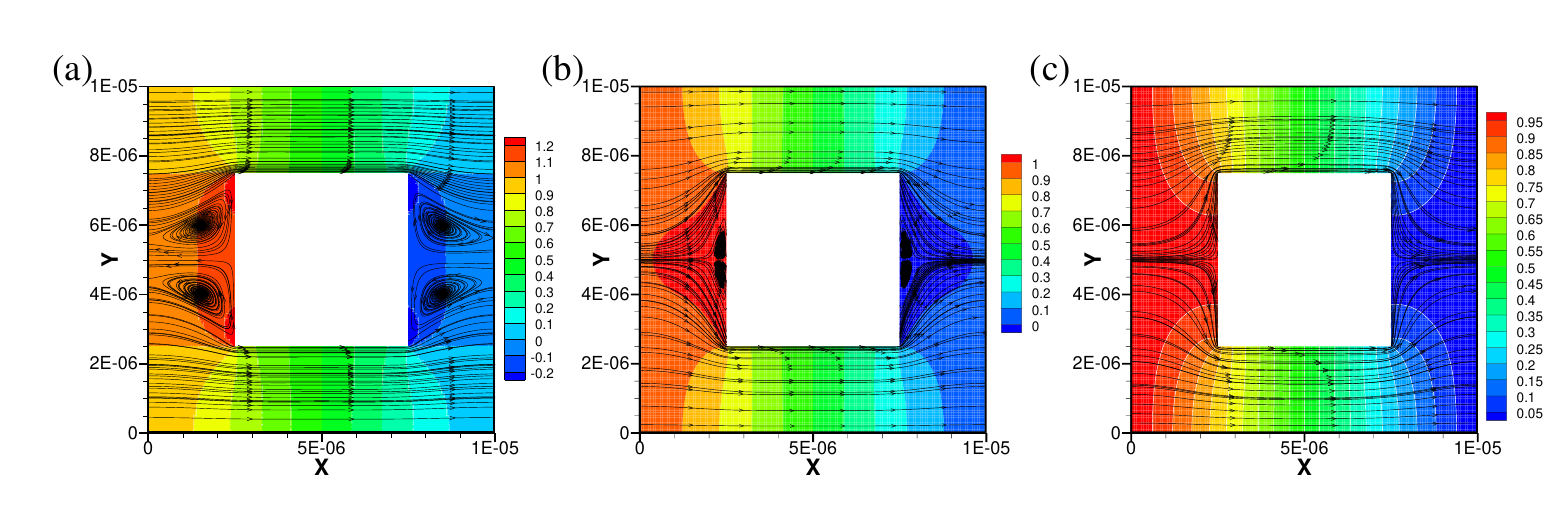}
     \caption{The normalized temperature contour $(T-T_c)/\Delta T$ and heat flux streamline in porous graphene (\cref{porous}) with different $T_0$, where $L=2d=10~\mu$m, $X$ and $Y$ are the coordinates. (a) $T_0=30~\text{K}$, (b) $T_0=100~\text{K}$, (c) $T_0=300~\text{K}$.}
     \label{porousL10um}
\end{figure}
\begin{figure}
     \centering
     \includegraphics[scale=1.1,viewport=50 150 500 300,clip=true]{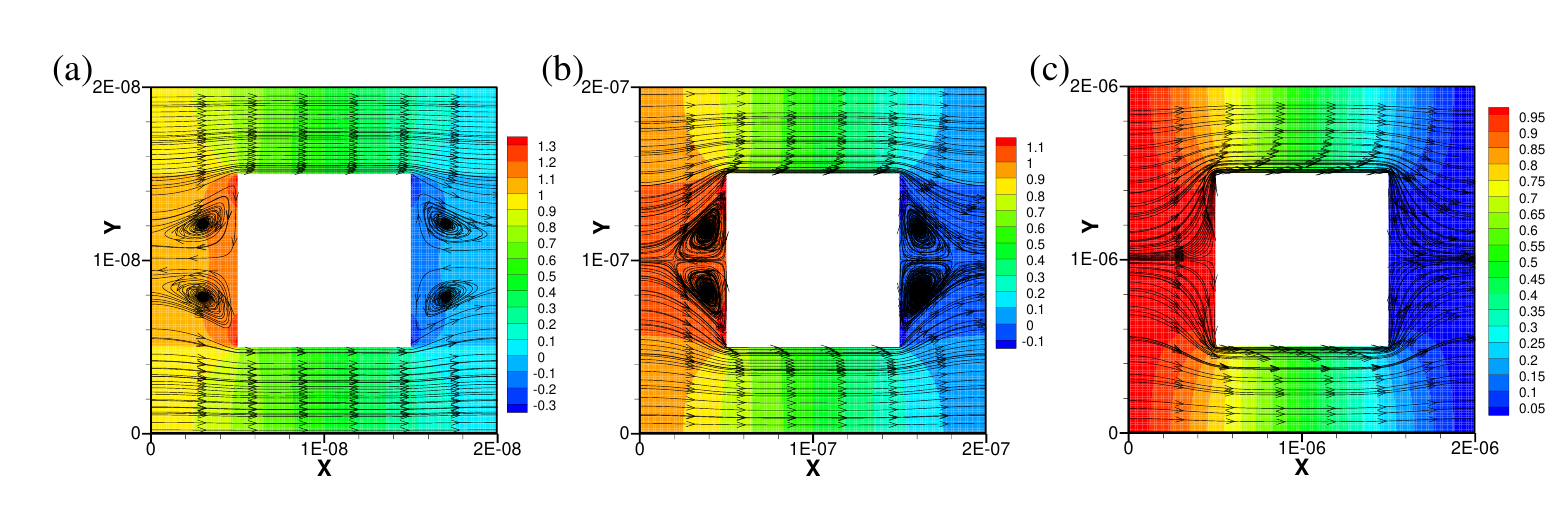}
     \caption{The normalized temperature contour $(T-T_c)/\Delta T$ and heat flux streamline in porous graphene (\cref{porous}) with different system sizes $L$, where $T_0 =300~ \text{K} $, $X$ and $Y$ are the coordinates. (a) $L=20$ nm, (b) $L=200$ nm, (c) $L=2~\mu $m.}
     \label{porousT300KL}
\end{figure}

To make the results more convincing, the porous graphene (isotopically pure) is taken as a specific example accounting for the phonon dispersion and polarization~\cite{PhysRevLett.95.096105,PhysRevB.79.155413,PhysRevB.93.235423}.
The detailed mathematical formulas of the phonon dispersion curves and scattering can refer to Ref~\cite{luo2019}.

First, the heat vortexes phenomena with different temperatures $T_0$ are predicted first for a given system size $L=10~\mu$m, as shown in~\cref{porousL10um}.
It can be observed that at $T_0 =30~\text{K}$, there are huge heat vortexes in front of the pore.
With the increase of $T_0$, the heat vortexes fade away gradually.
Because at low temperature, there are strong N scattering and weak R scattering in graphene~\cite{lee_hydrodynamic_2015,cepellotti_phonon_2015}.
As the temperature increases, the R scattering happens very frequently and dominates heat conduction, which recovers the Fourier's law of thermal conduction without heat vortexes.

Second, the heat vortexes phenomena with different system sizes $L$ are predicted at $T_0=300~\text{K}$.
From the results shown in~\cref{porousT300KL}(a)(b), it can be observed that as system size is smaller than or comparable to the phonon mean free path, the heat vortexes appear because boundary scattering (or ballistic phonon transport) dominates heat conduction and R scattering is weak in this length scale~\cite{RevModPhys.90.041002,xu_length-dependent_2014}.
As the system size increases, the R scattering increases so that the vortex sizes become small till zero at micron scale (\cref{porousT300KL}(c)).

In other words, the heat vortexes can appear in porous graphene at low temperature or nanostructures, where the R scattering is not sufficient~\cite{kaviany_2008,lee_hydrodynamic_2015,cepellotti_phonon_2015,RevModPhys.90.041002,xu_length-dependent_2014}.
And it is not easy to clearly distinguish the ballistic or hydrodynamic phonon transport in porous graphene only by the heat vortexes because the heat flux profiles in these regimes are very similar as shown in~\cref{porousT300KL}(a) and \cref{porousL10um}(a).

\subsection{Heat vortex in a 2D ribbon}

In previous studies~\cite{shang_heat_2020}, the heat vortexes in a graphene ribbon have been studied in the diffusive and hydrodynamic regimes with frequency independent relaxation time.
The heat vortexes appear in the hydrodynamic regime but disappear in the diffusive regime.
However, no comparison is made between the heat vortexes in the ballistic and hydrodynamic regimes.
And what will happen if considering the frequency dependent relaxation time and group velocity?

\subsubsection{Problem description}

\begin{figure}
 \centering
 \includegraphics[scale=0.30,viewport=100 130 800 450,clip=true]{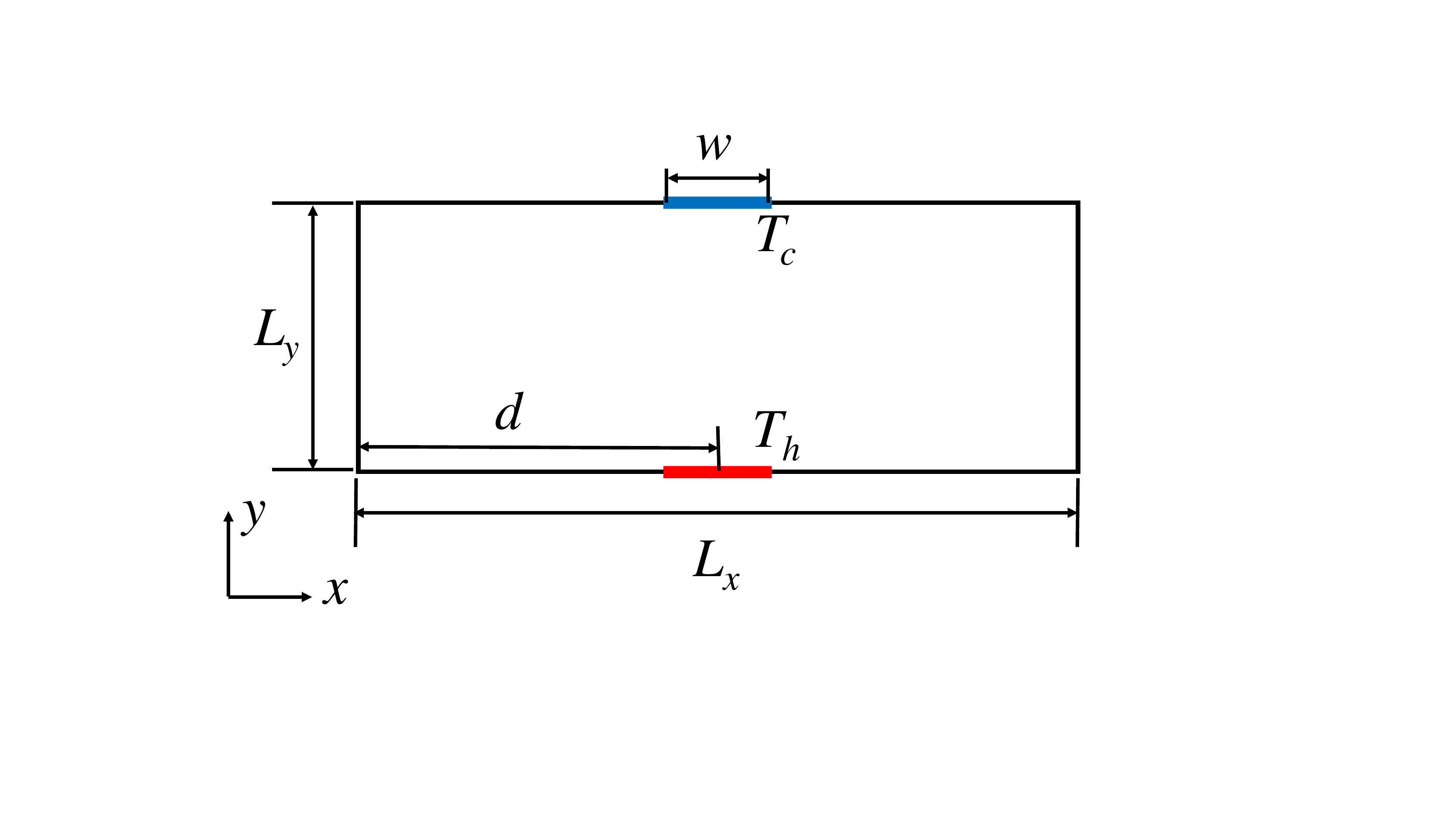}
 \caption{Schematic of heat conduction in a 2D ribbon, where the length and width of the geometry are $L_x$ and $L_y$, respectively. The heat source and heat sink are set at the bottom and top of the ribbon with width $w=0.4 L_y$, respectively. $d=2.5 L_y$ is the distance between the heat source (sink) and the left boundary of geometry. The temperatures of heat source and heat sink are $T_h$ and $T_c$, respectively, where $T_h=T_0 +\Delta T/2$, $T_c= T_0 -\Delta T/2$, $\Delta T/T_0=0.01$. So that the heat flows from the bottom to the top~\cite{shang_heat_2020}.}
 \label{ribbon}
\end{figure}

In order to answer above questions, the heat vortexes in a 2D ribbon are investigated in all regimes.
The schematic of 2D ribbon is shown in~\cref{ribbon}, where its side length and width are $L_x$ and $L_y$, respectively.
The heat source and heat sink are set at the bottom and top of the ribbon with width $w=0.4 L_y$, respectively.
$d=2.5 L_y$ is the distance between the heat source (sink) and the left boundary of geometry.
The temperatures of heat source and heat sink are $T_h$ and $T_c$, respectively, where $T_h=T_0 +\Delta T/2$, $T_c= T_0 -\Delta T/2$, $\Delta T/T_0=0.01$.
The initial temperature inside the domain is $T_0$.
The isothermal boundary conditions (Eq.~\eqref{eq:BC3}) are used for the heat source and sink.
For other boundaries, the diffusely reflecting boundary conditions (Eq.~\eqref{eq:BC1}) are implemented.

\begin{figure}
 \centering
 \includegraphics[scale=1.2,viewport=0 450 280 730,clip=true]{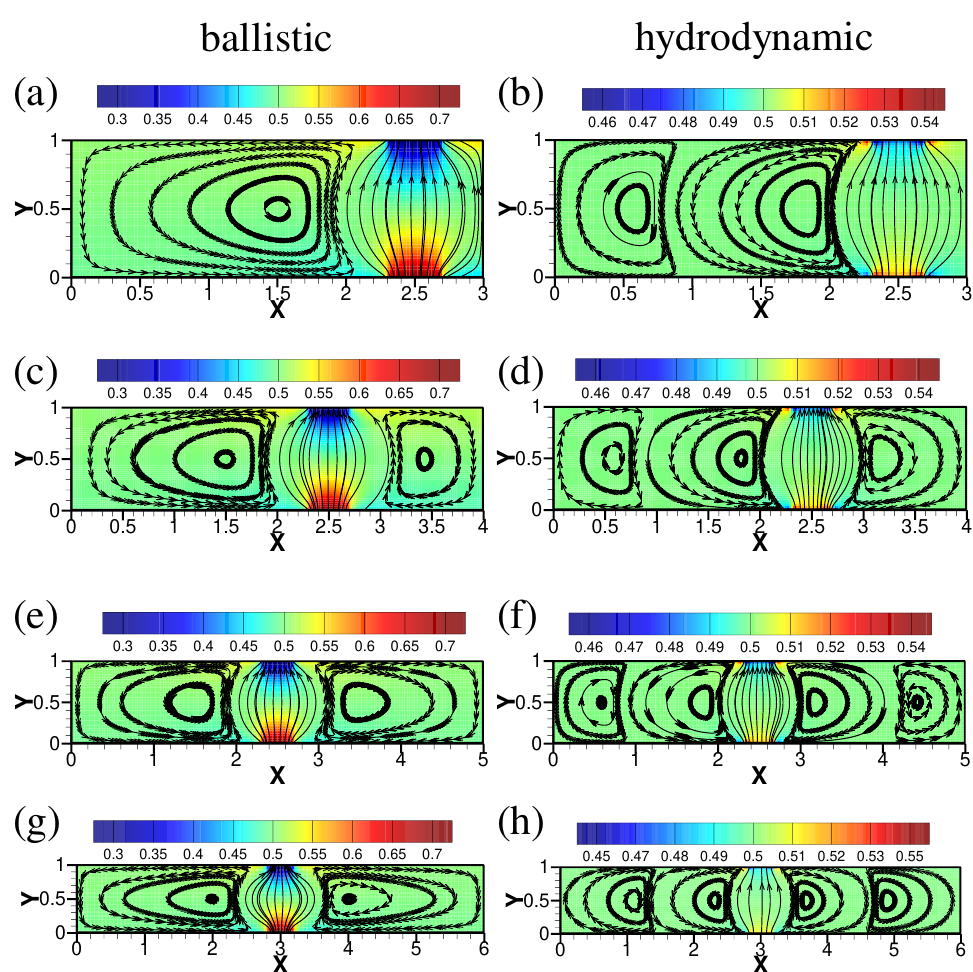}
 \caption{Heat vortexes in a ribbon (\cref{ribbon}) with different $L_x$ and fixed $L_y$ based on gray model~\cite{ChenG05Oxford,kaviany_2008}. It is not limited by a specific material so that all physical variables are set to be dimensionless, where $L_y=1.0$, $C=1,~|\bm{v}|=1,~T_0=1$, $X$ and $Y$ are the coordinates.
  Colored background is the normalized temperature field ($(T-T_c)/ \Delta T$) and black line is the heat flux streamline. (a)(c)(e)(g) Ballistic phonon transport ($\tau_N^{-1}=0$, $\tau_R^{-1}=0$). (b)(d)(f)(h) Hydrodynamic phonon transport ($\tau_N=0.01$, $\tau_R^{-1}=0$). (a)(b) $L_x=3.0$. (c)(d) $L_x=4.0$. (e)(f) $L_x=5.0$. (g)(h) $L_x=6.0$.}
 \label{viscous_length_BH}
\end{figure}
\begin{figure}
 \centering
 \includegraphics[scale=1.2,viewport=0 360 300 710,clip=true]{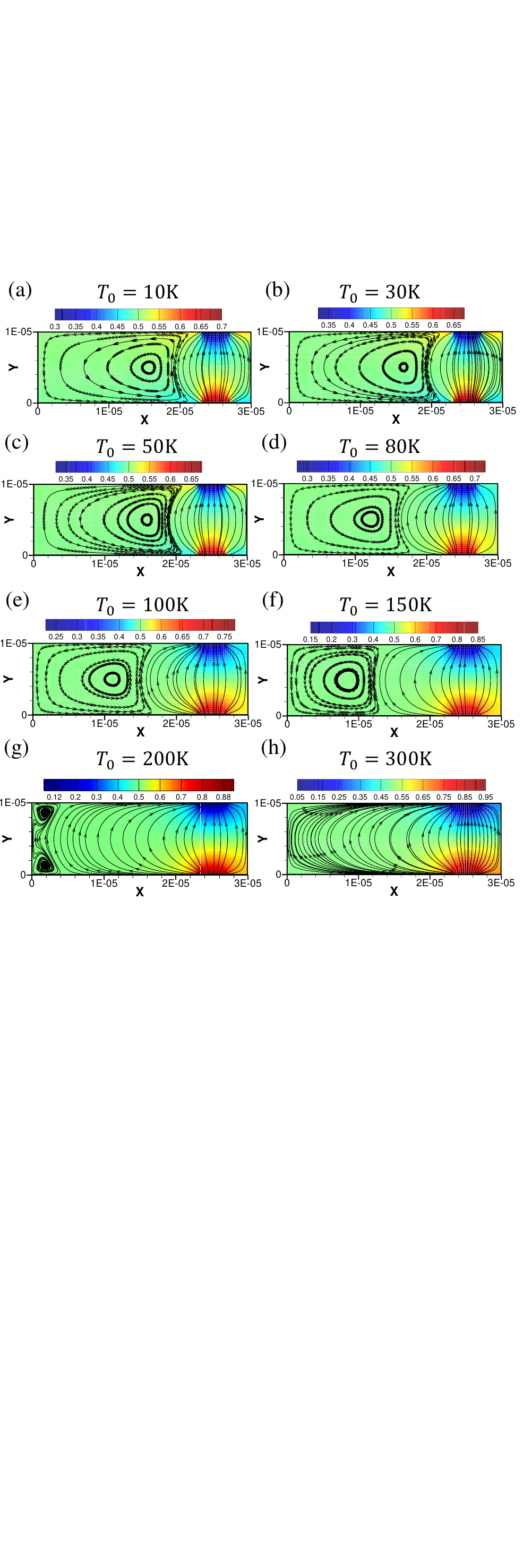}
 \caption{Heat vortexes in graphene ribbon (\cref{ribbon}) with different temperature $T_0$ and fixed system size $L_x=3 L_y =30 \mu$m. $X$ and $Y$ are the coordinates. Colored background is the normalized temperature field ($(T-T_c)/ \Delta T$) and black line is the heat flux streamline.}
 \label{viscous_length_GNR}
\end{figure}

\subsubsection{Gray model}

The gray model~\cite{ChenG05Oxford,kaviany_2008} is used here to investigate the ballistic ($\tau_R^{-1}=\tau_N^{-1}=0$) and hydrodynamic ($\tau_R^{-1}=0,~\tau_N=0.01$) phonon transport, where the width is fixed $L_y=1$ and the system length $L_x$ is changed.
Besides, $C=1$, $|\bm{v}|=1$, $T_0 =1$ and the thermal effects of R scattering is totally ignored.

Figure~\ref{viscous_length_BH} shows the heat vortexes in 2D ribbon of ballistic and hydrodynamic phonon transport with different system length $L_x$ ($L_x=3.0,~4.0,~5.0,~6.0$).
It can be observed that heat vortexes appear on the left side of heat source (sink) in the ballistic regime (\cref{viscous_length_BH}(a)(c)(e)(g)).
And as system length $L_x$ increases, the heat vortexes appears on the right side of heat source, too.
It indicates that the appearance of heat vortexes is related to the geometric configurations (or boundary conditions).
While in the hydrodynamic regime (\cref{viscous_length_BH}(b)(d)(f)(h)), there are heat vortexes, too.
Besides, the vortexes sizes are smaller than those in the ballistic regime with the same $L_x$.
As system length $L_x$ increases, the number of vortexes increases from one to two on the left side of heat source.

{\color{black}{Furthermore, it is also interesting to find that the biggest vortex size in the hydrodynamic regime is approximated as $L_y$ (\cref{viscous_length_BH}(b)(d)(f)), namely, the width of 2D ribbon.
While in the ballistic regime~\cite{MajumdarA93Film}, the biggest vortex size increases with the ribbon length $L_x$.
It may attribute to that in the ballistic regime, the phonon mean free path is much larger than the system size ($L_x$ or $L_y$).
In other words, the phonons can transport a very long distance without any change so that the information exchange in the whole physical space is very efficient.
However, in the phonon hydrodynamic regime~\cite{PhysRevB.102.094311,PhysRevB_SECOND_SOUND,lee_hydrodynamic_2015,cepellotti_phonon_2015,PhysRevB.99.085202}, frequent N scattering blocks the efficient information exchange in the physical space and the phonon distribution function approximately reaches a local equilibrium (Eq.~\eqref{eq:firstCEhydrodynamic}), not globally.
According to Eq.~\eqref{eq:steadyGKshang}, the boundary conditions or confined structures affect the vortex size, too.
So that the vortex size in the hydrodynamic regime does not increase with length $L_x$ endless but confined by the ribbon width $L_y$.
Maybe the vortex size in the ballistic or hydrodynamic limits can be theoretically derived based on the phonon BTE, which is beyond the scope of this work.}}

\subsubsection{Graphene ribbon}

The heat vortexes in graphene ribbon are studied and the frequency dependent relaxation time and group velocity are both considered~\cite{PhysRevLett.95.096105,PhysRevB.79.155413,PhysRevB.93.235423,luo2019}.
Different from last simulations without consideration of R scattering, the heat vortex in graphene ribbon is a competition result of boundary scattering, R scattering and N scattering.

According to previous studies~\cite{huberman_observation_2019,cepellotti_phonon_2015,luo2019}, the heat conduction in graphene ribbon with micron size goes through ballistic, hydrodynamic and diffusive regimes in turn as the temperature $T_0$ increases from extremely low to room temperature.
Besides, the ballistic and hydrodynamic regimes are very close~\cite{lee_hydrodynamic_2015,cepellotti_phonon_2015}, which both happen at low temperature.
So that in the following simulations, the system size of graphene ribbon is fixed, i.e., $L_x =3 L_y =30~\mu$m, and then the temperature $T_0$ is increased from $10$ K to $300$ K gradually.

As shown in~\cref{viscous_length_GNR}, it can be observed that as temperature increases, the vortex sizes changes slightly as $T_0 \leq 80$ K.
As $T_0 \geq 100$ K, the vortex size decreases gradually until zero in room temperature.
In addition, the number of heat vortexes is always one as $T_0 \leq 200$ K.
Namely, the number of heat vortexes does not change in graphene ribbon as phonon transport through ballistic and hydrodynamic regime, which is different from the results shown in~\cref{viscous_length_BH} without the consideration of R scattering.
That's because the R scattering impedes the formation of vortexes to some extent although it is weak at low temperature.

Therefore, it can be concluded that the heat vortexes can appear at low temperature, namely, as R scattering is not sufficient.
Besides, based on the present results in graphene ribbon (\cref{viscous_length_GNR}), it is very hard to distinguish the temperature ranges of ballistic or hydrodynamic phonon transport only by the heat vortexes.

\section{Conclusion}
\label{sec:conclusion}

In this work, the heat vortexes in two-dimensional porous or ribbon structures (\cref{porous,ribbon}) at steady state are investigated based on the phonon BTE under the Callaway model.
It is found that heat vortexes can appear in both ballistic and hydrodynamic regimes but disappear in the diffusive regime.
Furthermore, as long as there is not sufficient R scattering, the heat vortexes can appear in present simulations.
In addition, the vortex size in the hydrodynamic regime is smaller than that in the ballistic regime.
In ribbon structures, as the ribbon length increases, the vortex size increases endless in the ballistic regime but is confined by the ribbon width in the hydrodynamic regime.

In real two-dimensional materials (e.g., graphene), it is very hard to distinguish the ballistic and hydrodynamic phonon transport clearly by the heat vortexes at steady state.
Therefore, more work is needed to be done to distinguish the ballistic and hydrodynamic phonon transport clearly by the vortexes in the future, which may be inspired by recent studies of unsteady viscous electron flow~\cite{chandra2019a}.

\section*{Acknowledgments}

This study was supported by the National Key R$\&$D Program of China (No. 2018YFE0180900).
The authors acknowledge Jing-Tao L\"u, Manyu Shang and Rulei Guo for useful communications in phonon hydrodynamics.

\appendix

\section{Numerical solutions of the phonon BTE and boundary conditions}
\label{sec:callaway}

In this study, the stationary heat conduction is investigated so that the implicit discrete ordinate method is used to solve the stationary phonon BTE~\cite{ZHANG20191366,wangmr17callaway}.
The spatial space is discretized with $200 \times 200$ uniform cells for porous structures (\cref{porous}).
For ribbon structures (\cref{ribbon}), $300 \times 100$, $400 \times 200$ and $500 \times 100$ discretized spatial cells are used for different system lengths $L_x =3L_y$, $L_x =4L_y$ and $L_x =5L_y$, respectively.
For both structures, the first-order upwind scheme or van Leer limiter is used for spatial interpolations~\cite{ZHANG20191366}.
For the solid angle space, we set $\bm{s}=\left( \cos \theta, \sin \theta  \right)$, where $\theta \in [0, 2 \pi]$ is the polar angle.
Due to symmetry, the $\theta \in [0,\pi]$ is discretized with the $N_{\theta}$-point Gauss-Legendre quadrature~\cite{NicholasH13GaussL}.
In present simulations, we set $N_{\theta}=50$.

When the phonon dispersion and polarization are accounted~\cite{PhysRevLett.95.096105,PhysRevB.79.155413}, for each phonon acoustic branch in graphene (the optical branches are not accounted), the wave vector in single direction $k$ is discretized equally into $N_B$ discrete bands, i.e., $k_b=|\bm{K}|_{\text{m}} (2b-1)/(2N_B)$, where $b \in [1,N_B]$ is a positive integer, $0 \leq k \leq |\bm{K}|_{\text{m}}$, $|\bm{K}|_{\text{m}}= 1.5 \times 10^{10}\text{m}^{-1}$ is the maximum wave number.
Then the associated discretized phonon frequency, group velocity, phonon relaxation time can all be calculated.
The mid-point rule is used for the numerical integration of the frequency space.
In present simulations, we set $N_B=20$.

In addition, boundary conditions also play an indispensable role in numerical simulations~\cite{luo2019,ZHANG20191366}.
For the solution of the phonon BTE, the diffusely reflecting boundary condition, periodic boundary condition and isothermal boundary condition are presented as follows,
\begin{enumerate}
  \item Diffusely reflecting boundary condition assumes that the phonons reflected from the boundary are equal along each direction and the net heat flux across the boundary is zero, i.e.,
      \begin{align}
      e( \bm{x}_{b}, \bm{s})= \frac{1}{2}  \int_{ \bm{s}' \cdot \mathbf{n}_{b} < 0}{ e(\bm{x}_{b},\bm{s}')  \left| \bm{s}' \cdot  \mathbf{n}_{b}  \right| }d\Omega,
      \quad \bm{s} \cdot \mathbf{n}_{b} >0,
      \label{eq:BC1}
      \end{align}
      where $\mathbf{n}_{b}$ is the unit normal vector of the boundary pointing to the computational domain.
   \item In periodic boundary, when a phonon leaves the computational domain from one periodic boundary, another phonon with the same group velocity and frequency will enter the computational domain from the corresponding periodic boundary at the same time.
      Besides, the deviations from the local equilibrium states of the distribution functions of the two phonons are the same, i.e.,
      \begin{align}
      e(\bm{x}_{b_1} )- e^{eq}_{R} ({T}_{b_1}) = e(\bm{x}_{b_2} )- e^{eq}_{R} ({T}_{b_2}, ),
      \label{eq:BC2}
      \end{align}
      where $\bm{x}_{b_1}$, ${T}_{b_1}$ and $\bm{x}_{b_2}$, ${T}_{b_2}$ are the space vector and temperature of the two associated periodic boundaries $b_1$ and $b_2$, respectively.
  \item For isothermal boundary conditions, the distribution functions of phonons emitting from the boundaries with wall temperature $T_w$ are the equilibrium state of R scattering, i.e.,
      \begin{align}
      e(\bm{s}  ) =e^{eq}_R (T_w), \quad  \bm{s} \cdot \mathbf{n}_b >0 .
      \label{eq:BC3}
      \end{align}
\end{enumerate}

\bibliography{phonon}

\begin{thebibliography}{64}%
\makeatletter
\providecommand \@ifxundefined [1]{%
 \@ifx{#1\undefined}
}%
\providecommand \@ifnum [1]{%
 \ifnum #1\expandafter \@firstoftwo
 \else \expandafter \@secondoftwo
 \fi
}%
\providecommand \@ifx [1]{%
 \ifx #1\expandafter \@firstoftwo
 \else \expandafter \@secondoftwo
 \fi
}%
\providecommand \natexlab [1]{#1}%
\providecommand \enquote  [1]{``#1''}%
\providecommand \bibnamefont  [1]{#1}%
\providecommand \bibfnamefont [1]{#1}%
\providecommand \citenamefont [1]{#1}%
\providecommand \href@noop [0]{\@secondoftwo}%
\providecommand \href [0]{\begingroup \@sanitize@url \@href}%
\providecommand \@href[1]{\@@startlink{#1}\@@href}%
\providecommand \@@href[1]{\endgroup#1\@@endlink}%
\providecommand \@sanitize@url [0]{\catcode `\\12\catcode `\$12\catcode
  `\&12\catcode `\#12\catcode `\^12\catcode `\_12\catcode `\%12\relax}%
\providecommand \@@startlink[1]{}%
\providecommand \@@endlink[0]{}%
\providecommand \url  [0]{\begingroup\@sanitize@url \@url }%
\providecommand \@url [1]{\endgroup\@href {#1}{\urlprefix }}%
\providecommand \urlprefix  [0]{URL }%
\providecommand \Eprint [0]{\href }%
\providecommand \doibase [0]{http://dx.doi.org/}%
\providecommand \selectlanguage [0]{\@gobble}%
\providecommand \bibinfo  [0]{\@secondoftwo}%
\providecommand \bibfield  [0]{\@secondoftwo}%
\providecommand \translation [1]{[#1]}%
\providecommand \BibitemOpen [0]{}%
\providecommand \bibitemStop [0]{}%
\providecommand \bibitemNoStop [0]{.\EOS\space}%
\providecommand \EOS [0]{\spacefactor3000\relax}%
\providecommand \BibitemShut  [1]{\csname bibitem#1\endcsname}%
\let\auto@bib@innerbib\@empty
\bibitem [{\citenamefont {Kaviany}(2008)}]{kaviany_2008}%
  \BibitemOpen
  \bibfield  {author} {\bibinfo {author} {\bibfnamefont {M.}~\bibnamefont
  {Kaviany}},\ }\href {\doibase 10.1017/CBO9780511754586} {\emph {\bibinfo
  {title} {Heat transfer physics}}}\ (\bibinfo  {publisher} {Cambridge
  University Press},\ \bibinfo {year} {2008})\BibitemShut {NoStop}%
\bibitem [{\citenamefont {Chen}(2005)}]{ChenG05Oxford}%
  \BibitemOpen
  \bibfield  {author} {\bibinfo {author} {\bibfnamefont {G.}~\bibnamefont
  {Chen}},\ }\href
  {https://global.oup.com/ushe/product/nanoscale-energy-transport-and-conversion-9780195159424?cc=cn&lang=en&}
  {\emph {\bibinfo {title} {Nanoscale energy transport and conversion: a
  parallel treatment of electrons, molecules, phonons, and photons}}}\
  (\bibinfo  {publisher} {Oxford University Press},\ \bibinfo {year}
  {2005})\BibitemShut {NoStop}%
\bibitem [{\citenamefont {Liu}\ \emph {et~al.}(2014)\citenamefont {Liu},
  \citenamefont {Hänggi}, \citenamefont {Li}, \citenamefont {Ren},\ and\
  \citenamefont {Li}}]{liu_anomalous_2014}%
  \BibitemOpen
  \bibfield  {author} {\bibinfo {author} {\bibfnamefont {S.}~\bibnamefont
  {Liu}}, \bibinfo {author} {\bibfnamefont {P.}~\bibnamefont {Hänggi}},
  \bibinfo {author} {\bibfnamefont {N.}~\bibnamefont {Li}}, \bibinfo {author}
  {\bibfnamefont {J.}~\bibnamefont {Ren}}, \ and\ \bibinfo {author}
  {\bibfnamefont {B.}~\bibnamefont {Li}},\ }\href {\doibase
  10.1103/PhysRevLett.112.040601} {\bibfield  {journal} {\bibinfo  {journal}
  {Phys. Rev. Lett.}\ }\textbf {\bibinfo {volume} {112}},\ \bibinfo {pages}
  {040601} (\bibinfo {year} {2014})}\BibitemShut {NoStop}%
\bibitem [{\citenamefont {Joseph}\ and\ \citenamefont
  {Preziosi}(1989)}]{RevModPhysJoseph89}%
  \BibitemOpen
  \bibfield  {author} {\bibinfo {author} {\bibfnamefont {D.~D.}\ \bibnamefont
  {Joseph}}\ and\ \bibinfo {author} {\bibfnamefont {L.}~\bibnamefont
  {Preziosi}},\ }\href {\doibase 10.1103/RevModPhys.61.41} {\bibfield
  {journal} {\bibinfo  {journal} {Rev. Mod. Phys.}\ }\textbf {\bibinfo {volume}
  {61}},\ \bibinfo {pages} {41} (\bibinfo {year} {1989})}\BibitemShut {NoStop}%
\bibitem [{\citenamefont {Liao}(2020)}]{nanoscaleenergy2020}%
  \BibitemOpen
  \bibinfo {editor} {\bibfnamefont {B.}~\bibnamefont {Liao}},\ ed.,\ \href
  {\doibase 10.1088/978-0-7503-1738-2} {\emph {\bibinfo {title} {Nanoscale
  Energy Transport}}},\ 2053-2563\ (\bibinfo  {publisher} {IOP Publishing},\
  \bibinfo {year} {2020})\BibitemShut {NoStop}%
\bibitem [{\citenamefont {Li}\ \emph {et~al.}(2012)\citenamefont {Li},
  \citenamefont {Ren}, \citenamefont {Wang}, \citenamefont {Zhang},
  \citenamefont {H\"anggi},\ and\ \citenamefont {Li}}]{RevModPhysLibaowen}%
  \BibitemOpen
  \bibfield  {author} {\bibinfo {author} {\bibfnamefont {N.}~\bibnamefont
  {Li}}, \bibinfo {author} {\bibfnamefont {J.}~\bibnamefont {Ren}}, \bibinfo
  {author} {\bibfnamefont {L.}~\bibnamefont {Wang}}, \bibinfo {author}
  {\bibfnamefont {G.}~\bibnamefont {Zhang}}, \bibinfo {author} {\bibfnamefont
  {P.}~\bibnamefont {H\"anggi}}, \ and\ \bibinfo {author} {\bibfnamefont
  {B.}~\bibnamefont {Li}},\ }\href {\doibase 10.1103/RevModPhys.84.1045}
  {\bibfield  {journal} {\bibinfo  {journal} {Rev. Mod. Phys.}\ }\textbf
  {\bibinfo {volume} {84}},\ \bibinfo {pages} {1045} (\bibinfo {year}
  {2012})}\BibitemShut {NoStop}%
\bibitem [{\citenamefont {Gu}\ \emph {et~al.}(2018)\citenamefont {Gu},
  \citenamefont {Wei}, \citenamefont {Yin}, \citenamefont {Li},\ and\
  \citenamefont {Yang}}]{RevModPhys.90.041002}%
  \BibitemOpen
  \bibfield  {author} {\bibinfo {author} {\bibfnamefont {X.}~\bibnamefont
  {Gu}}, \bibinfo {author} {\bibfnamefont {Y.}~\bibnamefont {Wei}}, \bibinfo
  {author} {\bibfnamefont {X.}~\bibnamefont {Yin}}, \bibinfo {author}
  {\bibfnamefont {B.}~\bibnamefont {Li}}, \ and\ \bibinfo {author}
  {\bibfnamefont {R.}~\bibnamefont {Yang}},\ }\href {\doibase
  10.1103/RevModPhys.90.041002} {\bibfield  {journal} {\bibinfo  {journal}
  {Rev. Mod. Phys.}\ }\textbf {\bibinfo {volume} {90}},\ \bibinfo {pages}
  {041002} (\bibinfo {year} {2018})}\BibitemShut {NoStop}%
\bibitem [{\citenamefont {Chang}\ \emph {et~al.}(2008)\citenamefont {Chang},
  \citenamefont {Okawa}, \citenamefont {Garcia}, \citenamefont {Majumdar},\
  and\ \citenamefont {Zettl}}]{chang_breakdown_2008}%
  \BibitemOpen
  \bibfield  {author} {\bibinfo {author} {\bibfnamefont {C.~W.}\ \bibnamefont
  {Chang}}, \bibinfo {author} {\bibfnamefont {D.}~\bibnamefont {Okawa}},
  \bibinfo {author} {\bibfnamefont {H.}~\bibnamefont {Garcia}}, \bibinfo
  {author} {\bibfnamefont {A.}~\bibnamefont {Majumdar}}, \ and\ \bibinfo
  {author} {\bibfnamefont {A.}~\bibnamefont {Zettl}},\ }\href {\doibase
  10.1103/PhysRevLett.101.075903} {\bibfield  {journal} {\bibinfo  {journal}
  {Phys. Rev. Lett.}\ }\textbf {\bibinfo {volume} {101}},\ \bibinfo {pages}
  {075903} (\bibinfo {year} {2008})}\BibitemShut {NoStop}%
\bibitem [{\citenamefont {Hua}\ \emph {et~al.}(2019)\citenamefont {Hua},
  \citenamefont {Lindsay}, \citenamefont {Chen},\ and\ \citenamefont
  {Minnich}}]{PhysRevB.100.085203}%
  \BibitemOpen
  \bibfield  {author} {\bibinfo {author} {\bibfnamefont {C.}~\bibnamefont
  {Hua}}, \bibinfo {author} {\bibfnamefont {L.}~\bibnamefont {Lindsay}},
  \bibinfo {author} {\bibfnamefont {X.}~\bibnamefont {Chen}}, \ and\ \bibinfo
  {author} {\bibfnamefont {A.~J.}\ \bibnamefont {Minnich}},\ }\href {\doibase
  10.1103/PhysRevB.100.085203} {\bibfield  {journal} {\bibinfo  {journal}
  {Phys. Rev. B}\ }\textbf {\bibinfo {volume} {100}},\ \bibinfo {pages}
  {085203} (\bibinfo {year} {2019})}\BibitemShut {NoStop}%
\bibitem [{\citenamefont {Minnich}\ \emph {et~al.}(2011)\citenamefont
  {Minnich}, \citenamefont {Johnson}, \citenamefont {Schmidt}, \citenamefont
  {Esfarjani}, \citenamefont {Dresselhaus}, \citenamefont {Nelson},\ and\
  \citenamefont {Chen}}]{PhysRevLett.107.095901}%
  \BibitemOpen
  \bibfield  {author} {\bibinfo {author} {\bibfnamefont {A.~J.}\ \bibnamefont
  {Minnich}}, \bibinfo {author} {\bibfnamefont {J.~A.}\ \bibnamefont
  {Johnson}}, \bibinfo {author} {\bibfnamefont {A.~J.}\ \bibnamefont
  {Schmidt}}, \bibinfo {author} {\bibfnamefont {K.}~\bibnamefont {Esfarjani}},
  \bibinfo {author} {\bibfnamefont {M.~S.}\ \bibnamefont {Dresselhaus}},
  \bibinfo {author} {\bibfnamefont {K.~A.}\ \bibnamefont {Nelson}}, \ and\
  \bibinfo {author} {\bibfnamefont {G.}~\bibnamefont {Chen}},\ }\href {\doibase
  10.1103/PhysRevLett.107.095901} {\bibfield  {journal} {\bibinfo  {journal}
  {Phys. Rev. Lett.}\ }\textbf {\bibinfo {volume} {107}},\ \bibinfo {pages}
  {095901} (\bibinfo {year} {2011})}\BibitemShut {NoStop}%
\bibitem [{\citenamefont {Zhang}\ \emph {et~al.}(2020)\citenamefont {Zhang},
  \citenamefont {Ouyang}, \citenamefont {Cheng}, \citenamefont {Chen},
  \citenamefont {Li},\ and\ \citenamefont {Zhang}}]{ZHANG2020}%
  \BibitemOpen
  \bibfield  {author} {\bibinfo {author} {\bibfnamefont {Z.}~\bibnamefont
  {Zhang}}, \bibinfo {author} {\bibfnamefont {Y.}~\bibnamefont {Ouyang}},
  \bibinfo {author} {\bibfnamefont {Y.}~\bibnamefont {Cheng}}, \bibinfo
  {author} {\bibfnamefont {J.}~\bibnamefont {Chen}}, \bibinfo {author}
  {\bibfnamefont {N.}~\bibnamefont {Li}}, \ and\ \bibinfo {author}
  {\bibfnamefont {G.}~\bibnamefont {Zhang}},\ }\href {\doibase
  https://doi.org/10.1016/j.physrep.2020.03.001} {\bibfield  {journal}
  {\bibinfo  {journal} {Phys. Rep.}\ }\textbf {\bibinfo {volume} {860}},\
  \bibinfo {pages} {1 } (\bibinfo {year} {2020})}\BibitemShut {NoStop}%
\bibitem [{\citenamefont {Beardo}\ \emph {et~al.}(2020)\citenamefont {Beardo},
  \citenamefont {Hennessy}, \citenamefont {Sendra}, \citenamefont {Camacho},
  \citenamefont {Myers}, \citenamefont {Bafaluy},\ and\ \citenamefont
  {Alvarez}}]{PhysRevB.101.075303}%
  \BibitemOpen
  \bibfield  {author} {\bibinfo {author} {\bibfnamefont {A.}~\bibnamefont
  {Beardo}}, \bibinfo {author} {\bibfnamefont {M.~G.}\ \bibnamefont
  {Hennessy}}, \bibinfo {author} {\bibfnamefont {L.}~\bibnamefont {Sendra}},
  \bibinfo {author} {\bibfnamefont {J.}~\bibnamefont {Camacho}}, \bibinfo
  {author} {\bibfnamefont {T.~G.}\ \bibnamefont {Myers}}, \bibinfo {author}
  {\bibfnamefont {J.}~\bibnamefont {Bafaluy}}, \ and\ \bibinfo {author}
  {\bibfnamefont {F.~X.}\ \bibnamefont {Alvarez}},\ }\href {\doibase
  10.1103/PhysRevB.101.075303} {\bibfield  {journal} {\bibinfo  {journal}
  {Phys. Rev. B}\ }\textbf {\bibinfo {volume} {101}},\ \bibinfo {pages}
  {075303} (\bibinfo {year} {2020})}\BibitemShut {NoStop}%
\bibitem [{\citenamefont {de~Tomas}\ \emph {et~al.}(2014)\citenamefont
  {de~Tomas}, \citenamefont {Cantarero}, \citenamefont {Lopeandia},\ and\
  \citenamefont {Alvarez}}]{de_tomas_kinetic_2014}%
  \BibitemOpen
  \bibfield  {author} {\bibinfo {author} {\bibfnamefont {C.}~\bibnamefont
  {de~Tomas}}, \bibinfo {author} {\bibfnamefont {A.}~\bibnamefont {Cantarero}},
  \bibinfo {author} {\bibfnamefont {A.~F.}\ \bibnamefont {Lopeandia}}, \ and\
  \bibinfo {author} {\bibfnamefont {F.~X.}\ \bibnamefont {Alvarez}},\ }\href
  {\doibase 10.1063/1.4871672} {\bibfield  {journal} {\bibinfo  {journal} {J.
  Appl. Phys.}\ }\textbf {\bibinfo {volume} {115}},\ \bibinfo {pages} {164314}
  (\bibinfo {year} {2014})}\BibitemShut {NoStop}%
\bibitem [{\citenamefont {Guo}\ and\ \citenamefont
  {Wang}(2015)}]{WangMr15application}%
  \BibitemOpen
  \bibfield  {author} {\bibinfo {author} {\bibfnamefont {Y.}~\bibnamefont
  {Guo}}\ and\ \bibinfo {author} {\bibfnamefont {M.}~\bibnamefont {Wang}},\
  }\href {\doibase 10.1016/j.physrep.2015.07.003} {\bibfield  {journal}
  {\bibinfo  {journal} {Phys. Rep.}\ }\textbf {\bibinfo {volume} {595}},\
  \bibinfo {pages} {1 } (\bibinfo {year} {2015})}\BibitemShut {NoStop}%
\bibitem [{\citenamefont {Volz}\ and\ \citenamefont {Chen}(1999)}]{volz1999}%
  \BibitemOpen
  \bibfield  {author} {\bibinfo {author} {\bibfnamefont {S.~G.}\ \bibnamefont
  {Volz}}\ and\ \bibinfo {author} {\bibfnamefont {G.}~\bibnamefont {Chen}},\
  }\href {\doibase 10.1063/1.124914} {\bibfield  {journal} {\bibinfo  {journal}
  {Appl. Phys. Lett.}\ }\textbf {\bibinfo {volume} {75}},\ \bibinfo {pages}
  {2056} (\bibinfo {year} {1999})}\BibitemShut {NoStop}%
\bibitem [{\citenamefont {Majumdar}(1993)}]{MajumdarA93Film}%
  \BibitemOpen
  \bibfield  {author} {\bibinfo {author} {\bibfnamefont {A.}~\bibnamefont
  {Majumdar}},\ }\href {\doibase 10.1115/1.2910673} {\bibfield  {journal}
  {\bibinfo  {journal} {J. Heat Transfer}\ }\textbf {\bibinfo {volume} {115}},\
  \bibinfo {pages} {7} (\bibinfo {year} {1993})}\BibitemShut {NoStop}%
\bibitem [{\citenamefont {Beck}\ \emph {et~al.}(1974)\citenamefont {Beck},
  \citenamefont {Meier},\ and\ \citenamefont {Thellung}}]{beck1974}%
  \BibitemOpen
  \bibfield  {author} {\bibinfo {author} {\bibfnamefont {H.}~\bibnamefont
  {Beck}}, \bibinfo {author} {\bibfnamefont {P.~F.}\ \bibnamefont {Meier}}, \
  and\ \bibinfo {author} {\bibfnamefont {A.}~\bibnamefont {Thellung}},\ }\href
  {\doibase 10.1002/pssa.2210240102} {\bibfield  {journal} {\bibinfo  {journal}
  {Phys. Status Solidi A}\ }\textbf {\bibinfo {volume} {24}},\ \bibinfo {pages}
  {11} (\bibinfo {year} {1974})}\BibitemShut {NoStop}%
\bibitem [{\citenamefont {Narayanamurti}\ and\ \citenamefont
  {Dynes}(1972)}]{PhysRevLett.28.1461}%
  \BibitemOpen
  \bibfield  {author} {\bibinfo {author} {\bibfnamefont {V.}~\bibnamefont
  {Narayanamurti}}\ and\ \bibinfo {author} {\bibfnamefont {R.~C.}\ \bibnamefont
  {Dynes}},\ }\href {\doibase 10.1103/PhysRevLett.28.1461} {\bibfield
  {journal} {\bibinfo  {journal} {Phys. Rev. Lett.}\ }\textbf {\bibinfo
  {volume} {28}},\ \bibinfo {pages} {1461} (\bibinfo {year}
  {1972})}\BibitemShut {NoStop}%
\bibitem [{\citenamefont {McNelly}\ \emph {et~al.}(1970)\citenamefont
  {McNelly}, \citenamefont {Rogers}, \citenamefont {Channin}, \citenamefont
  {Rollefson}, \citenamefont {Goubau}, \citenamefont {Schmidt}, \citenamefont
  {Krumhansl},\ and\ \citenamefont {Pohl}}]{PhysRevLett_secondNaF}%
  \BibitemOpen
  \bibfield  {author} {\bibinfo {author} {\bibfnamefont {T.~F.}\ \bibnamefont
  {McNelly}}, \bibinfo {author} {\bibfnamefont {S.~J.}\ \bibnamefont {Rogers}},
  \bibinfo {author} {\bibfnamefont {D.~J.}\ \bibnamefont {Channin}}, \bibinfo
  {author} {\bibfnamefont {R.~J.}\ \bibnamefont {Rollefson}}, \bibinfo {author}
  {\bibfnamefont {W.~M.}\ \bibnamefont {Goubau}}, \bibinfo {author}
  {\bibfnamefont {G.~E.}\ \bibnamefont {Schmidt}}, \bibinfo {author}
  {\bibfnamefont {J.~A.}\ \bibnamefont {Krumhansl}}, \ and\ \bibinfo {author}
  {\bibfnamefont {R.~O.}\ \bibnamefont {Pohl}},\ }\href {\doibase
  10.1103/PhysRevLett.24.100} {\bibfield  {journal} {\bibinfo  {journal} {Phys.
  Rev. Lett.}\ }\textbf {\bibinfo {volume} {24}},\ \bibinfo {pages} {100}
  (\bibinfo {year} {1970})}\BibitemShut {NoStop}%
\bibitem [{\citenamefont {Jackson}\ \emph {et~al.}(1970)\citenamefont
  {Jackson}, \citenamefont {Walker},\ and\ \citenamefont
  {McNelly}}]{PhysRevLett_ssNaf}%
  \BibitemOpen
  \bibfield  {author} {\bibinfo {author} {\bibfnamefont {H.~E.}\ \bibnamefont
  {Jackson}}, \bibinfo {author} {\bibfnamefont {C.~T.}\ \bibnamefont {Walker}},
  \ and\ \bibinfo {author} {\bibfnamefont {T.~F.}\ \bibnamefont {McNelly}},\
  }\href {\doibase 10.1103/PhysRevLett.25.26} {\bibfield  {journal} {\bibinfo
  {journal} {Phys. Rev. Lett.}\ }\textbf {\bibinfo {volume} {25}},\ \bibinfo
  {pages} {26} (\bibinfo {year} {1970})}\BibitemShut {NoStop}%
\bibitem [{\citenamefont {Machida}\ \emph {et~al.}(2018)\citenamefont
  {Machida}, \citenamefont {Subedi}, \citenamefont {Akiba}, \citenamefont
  {Miyake}, \citenamefont {Tokunaga}, \citenamefont {Akahama}, \citenamefont
  {Izawa},\ and\ \citenamefont {Behnia}}]{machida2018}%
  \BibitemOpen
  \bibfield  {author} {\bibinfo {author} {\bibfnamefont {Y.}~\bibnamefont
  {Machida}}, \bibinfo {author} {\bibfnamefont {A.}~\bibnamefont {Subedi}},
  \bibinfo {author} {\bibfnamefont {K.}~\bibnamefont {Akiba}}, \bibinfo
  {author} {\bibfnamefont {A.}~\bibnamefont {Miyake}}, \bibinfo {author}
  {\bibfnamefont {M.}~\bibnamefont {Tokunaga}}, \bibinfo {author}
  {\bibfnamefont {Y.}~\bibnamefont {Akahama}}, \bibinfo {author} {\bibfnamefont
  {K.}~\bibnamefont {Izawa}}, \ and\ \bibinfo {author} {\bibfnamefont
  {K.}~\bibnamefont {Behnia}},\ }\href {\doibase 10.1126/sciadv.aat3374}
  {\bibfield  {journal} {\bibinfo  {journal} {Sci. Adv.}\ }\textbf {\bibinfo
  {volume} {4}},\ \bibinfo {pages} {eaat3374} (\bibinfo {year}
  {2018})}\BibitemShut {NoStop}%
\bibitem [{\citenamefont {Guyer}\ and\ \citenamefont
  {Krumhansl}(1966{\natexlab{a}})}]{PhysRev_GK}%
  \BibitemOpen
  \bibfield  {author} {\bibinfo {author} {\bibfnamefont {R.~A.}\ \bibnamefont
  {Guyer}}\ and\ \bibinfo {author} {\bibfnamefont {J.~A.}\ \bibnamefont
  {Krumhansl}},\ }\href {\doibase 10.1103/PhysRev.148.778} {\bibfield
  {journal} {\bibinfo  {journal} {Phys. Rev.}\ }\textbf {\bibinfo {volume}
  {148}},\ \bibinfo {pages} {778} (\bibinfo {year}
  {1966}{\natexlab{a}})}\BibitemShut {NoStop}%
\bibitem [{\citenamefont {Lee}\ \emph {et~al.}(2015)\citenamefont {Lee},
  \citenamefont {Broido}, \citenamefont {Esfarjani},\ and\ \citenamefont
  {Chen}}]{lee_hydrodynamic_2015}%
  \BibitemOpen
  \bibfield  {author} {\bibinfo {author} {\bibfnamefont {S.}~\bibnamefont
  {Lee}}, \bibinfo {author} {\bibfnamefont {D.}~\bibnamefont {Broido}},
  \bibinfo {author} {\bibfnamefont {K.}~\bibnamefont {Esfarjani}}, \ and\
  \bibinfo {author} {\bibfnamefont {G.}~\bibnamefont {Chen}},\ }\href {\doibase
  10.1038/ncomms7290} {\bibfield  {journal} {\bibinfo  {journal} {Nat.
  Commun.}\ }\textbf {\bibinfo {volume} {6}},\ \bibinfo {pages} {6290}
  (\bibinfo {year} {2015})}\BibitemShut {NoStop}%
\bibitem [{\citenamefont {Cepellotti}\ \emph {et~al.}(2015)\citenamefont
  {Cepellotti}, \citenamefont {Fugallo}, \citenamefont {Paulatto},
  \citenamefont {Lazzeri}, \citenamefont {Mauri},\ and\ \citenamefont
  {Marzari}}]{cepellotti_phonon_2015}%
  \BibitemOpen
  \bibfield  {author} {\bibinfo {author} {\bibfnamefont {A.}~\bibnamefont
  {Cepellotti}}, \bibinfo {author} {\bibfnamefont {G.}~\bibnamefont {Fugallo}},
  \bibinfo {author} {\bibfnamefont {L.}~\bibnamefont {Paulatto}}, \bibinfo
  {author} {\bibfnamefont {M.}~\bibnamefont {Lazzeri}}, \bibinfo {author}
  {\bibfnamefont {F.}~\bibnamefont {Mauri}}, \ and\ \bibinfo {author}
  {\bibfnamefont {N.}~\bibnamefont {Marzari}},\ }\href {\doibase
  10.1038/ncomms7400} {\bibfield  {journal} {\bibinfo  {journal} {Nat.
  Commun.}\ }\textbf {\bibinfo {volume} {6}} (\bibinfo {year} {2015}),\
  10.1038/ncomms7400}\BibitemShut {NoStop}%
\bibitem [{\citenamefont {Huberman}\ \emph {et~al.}(2019)\citenamefont
  {Huberman}, \citenamefont {Duncan}, \citenamefont {Chen}, \citenamefont
  {Song}, \citenamefont {Chiloyan}, \citenamefont {Ding}, \citenamefont
  {Maznev}, \citenamefont {Chen},\ and\ \citenamefont
  {Nelson}}]{huberman_observation_2019}%
  \BibitemOpen
  \bibfield  {author} {\bibinfo {author} {\bibfnamefont {S.}~\bibnamefont
  {Huberman}}, \bibinfo {author} {\bibfnamefont {R.~A.}\ \bibnamefont
  {Duncan}}, \bibinfo {author} {\bibfnamefont {K.}~\bibnamefont {Chen}},
  \bibinfo {author} {\bibfnamefont {B.}~\bibnamefont {Song}}, \bibinfo {author}
  {\bibfnamefont {V.}~\bibnamefont {Chiloyan}}, \bibinfo {author}
  {\bibfnamefont {Z.}~\bibnamefont {Ding}}, \bibinfo {author} {\bibfnamefont
  {A.~A.}\ \bibnamefont {Maznev}}, \bibinfo {author} {\bibfnamefont
  {G.}~\bibnamefont {Chen}}, \ and\ \bibinfo {author} {\bibfnamefont {K.~A.}\
  \bibnamefont {Nelson}},\ }\href {\doibase 10.1126/science.aav3548} {\bibfield
   {journal} {\bibinfo  {journal} {Science}\ ,\ \bibinfo {pages} {eaav3548}}
  (\bibinfo {year} {2019})}\BibitemShut {NoStop}%
\bibitem [{\citenamefont {Li}\ \emph {et~al.}(2003)\citenamefont {Li},
  \citenamefont {Wu}, \citenamefont {Kim}, \citenamefont {Shi}, \citenamefont
  {Yang},\ and\ \citenamefont {Majumdar}}]{li_thermal_2003}%
  \BibitemOpen
  \bibfield  {author} {\bibinfo {author} {\bibfnamefont {D.}~\bibnamefont
  {Li}}, \bibinfo {author} {\bibfnamefont {Y.}~\bibnamefont {Wu}}, \bibinfo
  {author} {\bibfnamefont {P.}~\bibnamefont {Kim}}, \bibinfo {author}
  {\bibfnamefont {L.}~\bibnamefont {Shi}}, \bibinfo {author} {\bibfnamefont
  {P.}~\bibnamefont {Yang}}, \ and\ \bibinfo {author} {\bibfnamefont
  {A.}~\bibnamefont {Majumdar}},\ }\href {\doibase 10.1063/1.1616981}
  {\bibfield  {journal} {\bibinfo  {journal} {Appl. Phys. Lett.}\ }\textbf
  {\bibinfo {volume} {83}},\ \bibinfo {pages} {2934} (\bibinfo {year}
  {2003})}\BibitemShut {NoStop}%
\bibitem [{\citenamefont {Ju}\ and\ \citenamefont
  {Goodson}(1999)}]{ju1999phonon}%
  \BibitemOpen
  \bibfield  {author} {\bibinfo {author} {\bibfnamefont {Y.}~\bibnamefont
  {Ju}}\ and\ \bibinfo {author} {\bibfnamefont {K.}~\bibnamefont {Goodson}},\
  }\href {\doibase 10.1063/1.123994} {\bibfield  {journal} {\bibinfo  {journal}
  {Appl. Phys. Lett.}\ }\textbf {\bibinfo {volume} {74}},\ \bibinfo {pages}
  {3005} (\bibinfo {year} {1999})}\BibitemShut {NoStop}%
\bibitem [{\citenamefont {Xu}\ \emph {et~al.}(2014)\citenamefont {Xu},
  \citenamefont {Pereira}, \citenamefont {Wang}, \citenamefont {Wu},
  \citenamefont {Zhang}, \citenamefont {Zhao}, \citenamefont {Bae},
  \citenamefont {Bui}, \citenamefont {Xie}, \citenamefont {Thong},
  \citenamefont {Hong}, \citenamefont {Loh}, \citenamefont {Donadio},
  \citenamefont {Li},\ and\ \citenamefont
  {{\"O}zyilmaz}}]{xu_length-dependent_2014}%
  \BibitemOpen
  \bibfield  {author} {\bibinfo {author} {\bibfnamefont {X.}~\bibnamefont
  {Xu}}, \bibinfo {author} {\bibfnamefont {L.~F.~C.}\ \bibnamefont {Pereira}},
  \bibinfo {author} {\bibfnamefont {Y.}~\bibnamefont {Wang}}, \bibinfo {author}
  {\bibfnamefont {J.}~\bibnamefont {Wu}}, \bibinfo {author} {\bibfnamefont
  {K.}~\bibnamefont {Zhang}}, \bibinfo {author} {\bibfnamefont
  {X.}~\bibnamefont {Zhao}}, \bibinfo {author} {\bibfnamefont {S.}~\bibnamefont
  {Bae}}, \bibinfo {author} {\bibfnamefont {C.~T.}\ \bibnamefont {Bui}},
  \bibinfo {author} {\bibfnamefont {R.}~\bibnamefont {Xie}}, \bibinfo {author}
  {\bibfnamefont {J.~T.~L.}\ \bibnamefont {Thong}}, \bibinfo {author}
  {\bibfnamefont {B.~H.}\ \bibnamefont {Hong}}, \bibinfo {author}
  {\bibfnamefont {K.~P.}\ \bibnamefont {Loh}}, \bibinfo {author} {\bibfnamefont
  {D.}~\bibnamefont {Donadio}}, \bibinfo {author} {\bibfnamefont
  {B.}~\bibnamefont {Li}}, \ and\ \bibinfo {author} {\bibfnamefont
  {B.}~\bibnamefont {{\"O}zyilmaz}},\ }\href {\doibase 10.1038/ncomms4689}
  {\bibfield  {journal} {\bibinfo  {journal} {Nat. Commun.}\ }\textbf {\bibinfo
  {volume} {5}},\ \bibinfo {pages} {3689} (\bibinfo {year} {2014})}\BibitemShut
  {NoStop}%
\bibitem [{\citenamefont {Gurzhi}(1968)}]{Gurzhi_1968}%
  \BibitemOpen
  \bibfield  {author} {\bibinfo {author} {\bibfnamefont {R.~N.}\ \bibnamefont
  {Gurzhi}},\ }\href {\doibase 10.1070/pu1968v011n02abeh003815} {\bibfield
  {journal} {\bibinfo  {journal} {Sov. Phys.-Usp.}\ }\textbf {\bibinfo {volume}
  {11}},\ \bibinfo {pages} {255} (\bibinfo {year} {1968})}\BibitemShut
  {NoStop}%
\bibitem [{\citenamefont {Hardy}\ and\ \citenamefont
  {Albers}(1974)}]{PhysRevB.10.3546}%
  \BibitemOpen
  \bibfield  {author} {\bibinfo {author} {\bibfnamefont {R.~J.}\ \bibnamefont
  {Hardy}}\ and\ \bibinfo {author} {\bibfnamefont {D.~L.}\ \bibnamefont
  {Albers}},\ }\href {\doibase 10.1103/PhysRevB.10.3546} {\bibfield  {journal}
  {\bibinfo  {journal} {Phys. Rev. B}\ }\textbf {\bibinfo {volume} {10}},\
  \bibinfo {pages} {3546} (\bibinfo {year} {1974})}\BibitemShut {NoStop}%
\bibitem [{\citenamefont {Enz}(1968)}]{ENZ1968114}%
  \BibitemOpen
  \bibfield  {author} {\bibinfo {author} {\bibfnamefont {C.~P.}\ \bibnamefont
  {Enz}},\ }\href {\doibase https://doi.org/10.1016/0003-4916(68)90305-9}
  {\bibfield  {journal} {\bibinfo  {journal} {Ann. Phys.}\ }\textbf {\bibinfo
  {volume} {46}},\ \bibinfo {pages} {114 } (\bibinfo {year}
  {1968})}\BibitemShut {NoStop}%
\bibitem [{\citenamefont {Ghosh}\ \emph {et~al.}(2020)\citenamefont {Ghosh},
  \citenamefont {Kusiak},\ and\ \citenamefont
  {Battaglia}}]{PhysRevB.102.094311}%
  \BibitemOpen
  \bibfield  {author} {\bibinfo {author} {\bibfnamefont {K.}~\bibnamefont
  {Ghosh}}, \bibinfo {author} {\bibfnamefont {A.}~\bibnamefont {Kusiak}}, \
  and\ \bibinfo {author} {\bibfnamefont {J.-L.}\ \bibnamefont {Battaglia}},\
  }\href {\doibase 10.1103/PhysRevB.102.094311} {\bibfield  {journal} {\bibinfo
   {journal} {Phys. Rev. B}\ }\textbf {\bibinfo {volume} {102}},\ \bibinfo
  {pages} {094311} (\bibinfo {year} {2020})}\BibitemShut {NoStop}%
\bibitem [{\citenamefont {Hardy}(1970)}]{PhysRevB_SECOND_SOUND}%
  \BibitemOpen
  \bibfield  {author} {\bibinfo {author} {\bibfnamefont {R.~J.}\ \bibnamefont
  {Hardy}},\ }\href {\doibase 10.1103/PhysRevB.2.1193} {\bibfield  {journal}
  {\bibinfo  {journal} {Phys. Rev. B}\ }\textbf {\bibinfo {volume} {2}},\
  \bibinfo {pages} {1193} (\bibinfo {year} {1970})}\BibitemShut {NoStop}%
\bibitem [{\citenamefont {Li}\ and\ \citenamefont
  {Lee}(2019)}]{PhysRevB.99.085202}%
  \BibitemOpen
  \bibfield  {author} {\bibinfo {author} {\bibfnamefont {X.}~\bibnamefont
  {Li}}\ and\ \bibinfo {author} {\bibfnamefont {S.}~\bibnamefont {Lee}},\
  }\href {\doibase 10.1103/PhysRevB.99.085202} {\bibfield  {journal} {\bibinfo
  {journal} {Phys. Rev. B}\ }\textbf {\bibinfo {volume} {99}},\ \bibinfo
  {pages} {085202} (\bibinfo {year} {2019})}\BibitemShut {NoStop}%
\bibitem [{\citenamefont {Martelli}\ \emph {et~al.}(2018)\citenamefont
  {Martelli}, \citenamefont {Jim\'enez}, \citenamefont {Continentino},
  \citenamefont {Baggio-Saitovitch},\ and\ \citenamefont
  {Behnia}}]{PhysRevLett_Strontium_Titanate}%
  \BibitemOpen
  \bibfield  {author} {\bibinfo {author} {\bibfnamefont {V.}~\bibnamefont
  {Martelli}}, \bibinfo {author} {\bibfnamefont {J.~L.}\ \bibnamefont
  {Jim\'enez}}, \bibinfo {author} {\bibfnamefont {M.}~\bibnamefont
  {Continentino}}, \bibinfo {author} {\bibfnamefont {E.}~\bibnamefont
  {Baggio-Saitovitch}}, \ and\ \bibinfo {author} {\bibfnamefont
  {K.}~\bibnamefont {Behnia}},\ }\href {\doibase
  10.1103/PhysRevLett.120.125901} {\bibfield  {journal} {\bibinfo  {journal}
  {Phys. Rev. Lett.}\ }\textbf {\bibinfo {volume} {120}},\ \bibinfo {pages}
  {125901} (\bibinfo {year} {2018})}\BibitemShut {NoStop}%
\bibitem [{\citenamefont {jr.}(2010)}]{prandtl_fluid}%
  \BibitemOpen
  \bibfield  {author} {\bibinfo {author} {\bibfnamefont {H.~O.}\ \bibnamefont
  {jr.}},\ }\href {\doibase 10.1007/978-1-4419-1564-1} {\emph {\bibinfo {title}
  {Prandtl-Essentials of Fluid Mechanics}}},\ \bibinfo {edition} {3rd}\ ed.,\
  \bibinfo {series} {Applied Mathematical Sciences}, Vol.\ \bibinfo {volume}
  {158}\ (\bibinfo  {publisher} {Springer-Verlag New York},\ \bibinfo {year}
  {2010})\BibitemShut {NoStop}%
\bibitem [{\citenamefont {Liu}\ \emph {et~al.}(2016)\citenamefont {Liu},
  \citenamefont {Wang}, \citenamefont {Yang},\ and\ \citenamefont
  {Duan}}]{liu_new_2016vortex}%
  \BibitemOpen
  \bibfield  {author} {\bibinfo {author} {\bibfnamefont {C.}~\bibnamefont
  {Liu}}, \bibinfo {author} {\bibfnamefont {Y.}~\bibnamefont {Wang}}, \bibinfo
  {author} {\bibfnamefont {Y.}~\bibnamefont {Yang}}, \ and\ \bibinfo {author}
  {\bibfnamefont {Z.}~\bibnamefont {Duan}},\ }\href {\doibase
  10.1007/s11433-016-0022-6} {\bibfield  {journal} {\bibinfo  {journal} {Sci.
  China Phys. Mech. Astron.}\ }\textbf {\bibinfo {volume} {59}},\ \bibinfo
  {pages} {684711} (\bibinfo {year} {2016})}\BibitemShut {NoStop}%
\bibitem [{\citenamefont {Guo}\ and\ \citenamefont {Shu}(2013)}]{GuoZl13LB}%
  \BibitemOpen
  \bibfield  {author} {\bibinfo {author} {\bibfnamefont {Z.}~\bibnamefont
  {Guo}}\ and\ \bibinfo {author} {\bibfnamefont {C.}~\bibnamefont {Shu}},\
  }\href {http://www.worldscientific.com/worldscibooks/10.1142/8806} {\emph
  {\bibinfo {title} {Lattice {B}oltzmann method and its applications in
  engineering}}},\ Vol.~\bibinfo {volume} {3}\ (\bibinfo  {publisher} {World
  Scientific},\ \bibinfo {year} {2013})\BibitemShut {NoStop}%
\bibitem [{\citenamefont {Scuracchio}\ \emph {et~al.}(2019)\citenamefont
  {Scuracchio}, \citenamefont {Michel},\ and\ \citenamefont
  {Peeters}}]{PhysRevB.99.144303}%
  \BibitemOpen
  \bibfield  {author} {\bibinfo {author} {\bibfnamefont {P.}~\bibnamefont
  {Scuracchio}}, \bibinfo {author} {\bibfnamefont {K.~H.}\ \bibnamefont
  {Michel}}, \ and\ \bibinfo {author} {\bibfnamefont {F.~M.}\ \bibnamefont
  {Peeters}},\ }\href {\doibase 10.1103/PhysRevB.99.144303} {\bibfield
  {journal} {\bibinfo  {journal} {Phys. Rev. B}\ }\textbf {\bibinfo {volume}
  {99}},\ \bibinfo {pages} {144303} (\bibinfo {year} {2019})}\BibitemShut
  {NoStop}%
\bibitem [{\citenamefont {Ding}\ \emph {et~al.}(2018)\citenamefont {Ding},
  \citenamefont {Zhou}, \citenamefont {Song}, \citenamefont {Chiloyan},
  \citenamefont {Li}, \citenamefont {Liu},\ and\ \citenamefont
  {Chen}}]{nanoletterchengang_2018}%
  \BibitemOpen
  \bibfield  {author} {\bibinfo {author} {\bibfnamefont {Z.}~\bibnamefont
  {Ding}}, \bibinfo {author} {\bibfnamefont {J.}~\bibnamefont {Zhou}}, \bibinfo
  {author} {\bibfnamefont {B.}~\bibnamefont {Song}}, \bibinfo {author}
  {\bibfnamefont {V.}~\bibnamefont {Chiloyan}}, \bibinfo {author}
  {\bibfnamefont {M.}~\bibnamefont {Li}}, \bibinfo {author} {\bibfnamefont
  {T.-H.}\ \bibnamefont {Liu}}, \ and\ \bibinfo {author} {\bibfnamefont
  {G.}~\bibnamefont {Chen}},\ }\href {\doibase 10.1021/acs.nanolett.7b04932}
  {\bibfield  {journal} {\bibinfo  {journal} {Nano Lett.}\ }\textbf {\bibinfo
  {volume} {18}},\ \bibinfo {pages} {638} (\bibinfo {year} {2018})},\ \bibinfo
  {note} {pMID: 29236507}\BibitemShut {NoStop}%
\bibitem [{\citenamefont {Dreyer}\ and\ \citenamefont
  {Struchtrup}(1993)}]{Dreyer1993}%
  \BibitemOpen
  \bibfield  {author} {\bibinfo {author} {\bibfnamefont {W.}~\bibnamefont
  {Dreyer}}\ and\ \bibinfo {author} {\bibfnamefont {H.}~\bibnamefont
  {Struchtrup}},\ }\href {\doibase 10.1007/BF01135371} {\bibfield  {journal}
  {\bibinfo  {journal} {CONTINUUM. MECH. THERM.}\ }\textbf {\bibinfo {volume}
  {5}},\ \bibinfo {pages} {3} (\bibinfo {year} {1993})}\BibitemShut {NoStop}%
\bibitem [{\citenamefont {Machida}\ \emph {et~al.}(2020)\citenamefont
  {Machida}, \citenamefont {Matsumoto}, \citenamefont {Isono},\ and\
  \citenamefont {Behnia}}]{machida2020}%
  \BibitemOpen
  \bibfield  {author} {\bibinfo {author} {\bibfnamefont {Y.}~\bibnamefont
  {Machida}}, \bibinfo {author} {\bibfnamefont {N.}~\bibnamefont {Matsumoto}},
  \bibinfo {author} {\bibfnamefont {T.}~\bibnamefont {Isono}}, \ and\ \bibinfo
  {author} {\bibfnamefont {K.}~\bibnamefont {Behnia}},\ }\href {\doibase
  10.1126/science.aaz8043} {\bibfield  {journal} {\bibinfo  {journal}
  {Science}\ }\textbf {\bibinfo {volume} {367}},\ \bibinfo {pages} {309}
  (\bibinfo {year} {2020})}\BibitemShut {NoStop}%
\bibitem [{\citenamefont {Ackerman}\ \emph {et~al.}(1966)\citenamefont
  {Ackerman}, \citenamefont {Bertman}, \citenamefont {Fairbank},\ and\
  \citenamefont {Guyer}}]{PhysRevLett.16.789}%
  \BibitemOpen
  \bibfield  {author} {\bibinfo {author} {\bibfnamefont {C.~C.}\ \bibnamefont
  {Ackerman}}, \bibinfo {author} {\bibfnamefont {B.}~\bibnamefont {Bertman}},
  \bibinfo {author} {\bibfnamefont {H.~A.}\ \bibnamefont {Fairbank}}, \ and\
  \bibinfo {author} {\bibfnamefont {R.~A.}\ \bibnamefont {Guyer}},\ }\href
  {\doibase 10.1103/PhysRevLett.16.789} {\bibfield  {journal} {\bibinfo
  {journal} {Phys. Rev. Lett.}\ }\textbf {\bibinfo {volume} {16}},\ \bibinfo
  {pages} {789} (\bibinfo {year} {1966})}\BibitemShut {NoStop}%
\bibitem [{\citenamefont {Nie}\ and\ \citenamefont
  {Cao}(2020)}]{nie2020thermal}%
  \BibitemOpen
  \bibfield  {author} {\bibinfo {author} {\bibfnamefont {B.-D.}\ \bibnamefont
  {Nie}}\ and\ \bibinfo {author} {\bibfnamefont {B.-Y.}\ \bibnamefont {Cao}},\
  }\href {\doibase 10.1080/15567265.2020.1755399} {\bibfield  {journal}
  {\bibinfo  {journal} {Nanoscale and Microscale Thermophysical Engineering}\
  ,\ \bibinfo {pages} {1}} (\bibinfo {year} {2020})}\BibitemShut {NoStop}%
\bibitem [{\citenamefont {Guyer}\ and\ \citenamefont
  {Krumhansl}(1966{\natexlab{b}})}]{PhysRev.148.766}%
  \BibitemOpen
  \bibfield  {author} {\bibinfo {author} {\bibfnamefont {R.~A.}\ \bibnamefont
  {Guyer}}\ and\ \bibinfo {author} {\bibfnamefont {J.~A.}\ \bibnamefont
  {Krumhansl}},\ }\href {\doibase 10.1103/PhysRev.148.766} {\bibfield
  {journal} {\bibinfo  {journal} {Phys. Rev.}\ }\textbf {\bibinfo {volume}
  {148}},\ \bibinfo {pages} {766} (\bibinfo {year}
  {1966}{\natexlab{b}})}\BibitemShut {NoStop}%
\bibitem [{\citenamefont {Shang}\ \emph {et~al.}(2020)\citenamefont {Shang},
  \citenamefont {Zhang}, \citenamefont {Guo},\ and\ \citenamefont
  {Lü}}]{shang_heat_2020}%
  \BibitemOpen
  \bibfield  {author} {\bibinfo {author} {\bibfnamefont {M.-Y.}\ \bibnamefont
  {Shang}}, \bibinfo {author} {\bibfnamefont {C.}~\bibnamefont {Zhang}},
  \bibinfo {author} {\bibfnamefont {Z.}~\bibnamefont {Guo}}, \ and\ \bibinfo
  {author} {\bibfnamefont {J.-T.}\ \bibnamefont {Lü}},\ }\href {\doibase
  10.1038/s41598-020-65221-8} {\bibfield  {journal} {\bibinfo  {journal} {Sci.
  Rep.}\ }\textbf {\bibinfo {volume} {10}},\ \bibinfo {pages} {8272} (\bibinfo
  {year} {2020})}\BibitemShut {NoStop}%
\bibitem [{\citenamefont {Simoncelli}\ \emph {et~al.}(2020)\citenamefont
  {Simoncelli}, \citenamefont {Marzari},\ and\ \citenamefont
  {Cepellotti}}]{PhysRevX.10.011019}%
  \BibitemOpen
  \bibfield  {author} {\bibinfo {author} {\bibfnamefont {M.}~\bibnamefont
  {Simoncelli}}, \bibinfo {author} {\bibfnamefont {N.}~\bibnamefont {Marzari}},
  \ and\ \bibinfo {author} {\bibfnamefont {A.}~\bibnamefont {Cepellotti}},\
  }\href {\doibase 10.1103/PhysRevX.10.011019} {\bibfield  {journal} {\bibinfo
  {journal} {Phys. Rev. X}\ }\textbf {\bibinfo {volume} {10}},\ \bibinfo
  {pages} {011019} (\bibinfo {year} {2020})}\BibitemShut {NoStop}%
\bibitem [{\citenamefont {Levitov}\ and\ \citenamefont
  {Falkovich}(2016)}]{levitov_electron_2016}%
  \BibitemOpen
  \bibfield  {author} {\bibinfo {author} {\bibfnamefont {L.}~\bibnamefont
  {Levitov}}\ and\ \bibinfo {author} {\bibfnamefont {G.}~\bibnamefont
  {Falkovich}},\ }\href {\doibase 10.1038/nphys3667} {\bibfield  {journal}
  {\bibinfo  {journal} {Nat. Phys.}\ }\textbf {\bibinfo {volume} {12}},\
  \bibinfo {pages} {672} (\bibinfo {year} {2016})}\BibitemShut {NoStop}%
\bibitem [{\citenamefont {Bandurin}\ \emph {et~al.}(2016)\citenamefont
  {Bandurin}, \citenamefont {Torre}, \citenamefont {Kumar}, \citenamefont
  {Shalom}, \citenamefont {Tomadin}, \citenamefont {Principi}, \citenamefont
  {Auton}, \citenamefont {Khestanova}, \citenamefont {Novoselov}, \citenamefont
  {Grigorieva}, \citenamefont {Ponomarenko}, \citenamefont {Geim},\ and\
  \citenamefont {Polini}}]{bandurin2016b}%
  \BibitemOpen
  \bibfield  {author} {\bibinfo {author} {\bibfnamefont {D.~A.}\ \bibnamefont
  {Bandurin}}, \bibinfo {author} {\bibfnamefont {I.}~\bibnamefont {Torre}},
  \bibinfo {author} {\bibfnamefont {R.~K.}\ \bibnamefont {Kumar}}, \bibinfo
  {author} {\bibfnamefont {M.~B.}\ \bibnamefont {Shalom}}, \bibinfo {author}
  {\bibfnamefont {A.}~\bibnamefont {Tomadin}}, \bibinfo {author} {\bibfnamefont
  {A.}~\bibnamefont {Principi}}, \bibinfo {author} {\bibfnamefont {G.~H.}\
  \bibnamefont {Auton}}, \bibinfo {author} {\bibfnamefont {E.}~\bibnamefont
  {Khestanova}}, \bibinfo {author} {\bibfnamefont {K.~S.}\ \bibnamefont
  {Novoselov}}, \bibinfo {author} {\bibfnamefont {I.~V.}\ \bibnamefont
  {Grigorieva}}, \bibinfo {author} {\bibfnamefont {L.~A.}\ \bibnamefont
  {Ponomarenko}}, \bibinfo {author} {\bibfnamefont {A.~K.}\ \bibnamefont
  {Geim}}, \ and\ \bibinfo {author} {\bibfnamefont {M.}~\bibnamefont
  {Polini}},\ }\href {\doibase 10.1126/science.aad0201} {\bibfield  {journal}
  {\bibinfo  {journal} {Science}\ }\textbf {\bibinfo {volume} {351}},\ \bibinfo
  {pages} {1055} (\bibinfo {year} {2016})}\BibitemShut {NoStop}%
\bibitem [{\citenamefont {Chandra}\ \emph {et~al.}(2019)\citenamefont
  {Chandra}, \citenamefont {Kataria}, \citenamefont {Sahdev},\ and\
  \citenamefont {Sundararaman}}]{chandra2019a}%
  \BibitemOpen
  \bibfield  {author} {\bibinfo {author} {\bibfnamefont {M.}~\bibnamefont
  {Chandra}}, \bibinfo {author} {\bibfnamefont {G.}~\bibnamefont {Kataria}},
  \bibinfo {author} {\bibfnamefont {D.}~\bibnamefont {Sahdev}}, \ and\ \bibinfo
  {author} {\bibfnamefont {R.}~\bibnamefont {Sundararaman}},\ }\href {\doibase
  10.1103/PhysRevB.99.165409} {\bibfield  {journal} {\bibinfo  {journal} {Phys.
  Rev. B}\ }\textbf {\bibinfo {volume} {99}},\ \bibinfo {pages} {165409}
  (\bibinfo {year} {2019})}\BibitemShut {NoStop}%
\bibitem [{\citenamefont {Danz}\ and\ \citenamefont
  {Narozhny}(2020)}]{Danz_2020}%
  \BibitemOpen
  \bibfield  {author} {\bibinfo {author} {\bibfnamefont {S.}~\bibnamefont
  {Danz}}\ and\ \bibinfo {author} {\bibfnamefont {B.~N.}\ \bibnamefont
  {Narozhny}},\ }\href {\doibase 10.1088/2053-1583/ab7bfa} {\bibfield
  {journal} {\bibinfo  {journal} {2D Mater.}\ }\textbf {\bibinfo {volume}
  {7}},\ \bibinfo {pages} {035001} (\bibinfo {year} {2020})}\BibitemShut
  {NoStop}%
\bibitem [{\citenamefont {Callaway}(1959)}]{PhysRev_callaway}%
  \BibitemOpen
  \bibfield  {author} {\bibinfo {author} {\bibfnamefont {J.}~\bibnamefont
  {Callaway}},\ }\href {\doibase 10.1103/PhysRev.113.1046} {\bibfield
  {journal} {\bibinfo  {journal} {Phys. Rev.}\ }\textbf {\bibinfo {volume}
  {113}},\ \bibinfo {pages} {1046} (\bibinfo {year} {1959})}\BibitemShut
  {NoStop}%
\bibitem [{\citenamefont {Guo}\ and\ \citenamefont
  {Wang}(2017)}]{wangmr17callaway}%
  \BibitemOpen
  \bibfield  {author} {\bibinfo {author} {\bibfnamefont {Y.}~\bibnamefont
  {Guo}}\ and\ \bibinfo {author} {\bibfnamefont {M.}~\bibnamefont {Wang}},\
  }\href {\doibase 10.1103/PhysRevB.96.134312} {\bibfield  {journal} {\bibinfo
  {journal} {Phys. Rev. B}\ }\textbf {\bibinfo {volume} {96}},\ \bibinfo
  {pages} {134312} (\bibinfo {year} {2017})}\BibitemShut {NoStop}%
\bibitem [{\citenamefont {Luo}\ \emph {et~al.}(2019)\citenamefont {Luo},
  \citenamefont {Guo}, \citenamefont {Wang},\ and\ \citenamefont
  {Yi}}]{luo2019}%
  \BibitemOpen
  \bibfield  {author} {\bibinfo {author} {\bibfnamefont {X.-P.}\ \bibnamefont
  {Luo}}, \bibinfo {author} {\bibfnamefont {Y.-Y.}\ \bibnamefont {Guo}},
  \bibinfo {author} {\bibfnamefont {M.-R.}\ \bibnamefont {Wang}}, \ and\
  \bibinfo {author} {\bibfnamefont {H.-L.}\ \bibnamefont {Yi}},\ }\href
  {\doibase 10.1103/PhysRevB.100.155401} {\bibfield  {journal} {\bibinfo
  {journal} {Phys. Rev. B}\ }\textbf {\bibinfo {volume} {100}},\ \bibinfo
  {pages} {155401} (\bibinfo {year} {2019})}\BibitemShut {NoStop}%
\bibitem [{\citenamefont {Mingo}\ and\ \citenamefont
  {Broido}(2005)}]{PhysRevLett.95.096105}%
  \BibitemOpen
  \bibfield  {author} {\bibinfo {author} {\bibfnamefont {N.}~\bibnamefont
  {Mingo}}\ and\ \bibinfo {author} {\bibfnamefont {D.~A.}\ \bibnamefont
  {Broido}},\ }\href {\doibase 10.1103/PhysRevLett.95.096105} {\bibfield
  {journal} {\bibinfo  {journal} {Phys. Rev. Lett.}\ }\textbf {\bibinfo
  {volume} {95}},\ \bibinfo {pages} {096105} (\bibinfo {year}
  {2005})}\BibitemShut {NoStop}%
\bibitem [{\citenamefont {Nika}\ \emph {et~al.}(2009)\citenamefont {Nika},
  \citenamefont {Pokatilov}, \citenamefont {Askerov},\ and\ \citenamefont
  {Balandin}}]{PhysRevB.79.155413}%
  \BibitemOpen
  \bibfield  {author} {\bibinfo {author} {\bibfnamefont {D.~L.}\ \bibnamefont
  {Nika}}, \bibinfo {author} {\bibfnamefont {E.~P.}\ \bibnamefont {Pokatilov}},
  \bibinfo {author} {\bibfnamefont {A.~S.}\ \bibnamefont {Askerov}}, \ and\
  \bibinfo {author} {\bibfnamefont {A.~A.}\ \bibnamefont {Balandin}},\ }\href
  {\doibase 10.1103/PhysRevB.79.155413} {\bibfield  {journal} {\bibinfo
  {journal} {Phys. Rev. B}\ }\textbf {\bibinfo {volume} {79}},\ \bibinfo
  {pages} {155413} (\bibinfo {year} {2009})}\BibitemShut {NoStop}%
\bibitem [{\citenamefont {Yang}\ and\ \citenamefont
  {Chen}(2003)}]{PhysRevB.67.195311}%
  \BibitemOpen
  \bibfield  {author} {\bibinfo {author} {\bibfnamefont {B.}~\bibnamefont
  {Yang}}\ and\ \bibinfo {author} {\bibfnamefont {G.}~\bibnamefont {Chen}},\
  }\href {\doibase 10.1103/PhysRevB.67.195311} {\bibfield  {journal} {\bibinfo
  {journal} {Phys. Rev. B}\ }\textbf {\bibinfo {volume} {67}},\ \bibinfo
  {pages} {195311} (\bibinfo {year} {2003})}\BibitemShut {NoStop}%
\bibitem [{\citenamefont {Guo}\ and\ \citenamefont {Xu}(2016)}]{GuoZl16DUGKS}%
  \BibitemOpen
  \bibfield  {author} {\bibinfo {author} {\bibfnamefont {Z.}~\bibnamefont
  {Guo}}\ and\ \bibinfo {author} {\bibfnamefont {K.}~\bibnamefont {Xu}},\
  }\href {\doibase 10.1016/j.ijheatmasstransfer.2016.06.088} {\bibfield
  {journal} {\bibinfo  {journal} {Int. J. Heat Mass Transfer}\ }\textbf
  {\bibinfo {volume} {102}},\ \bibinfo {pages} {944 } (\bibinfo {year}
  {2016})}\BibitemShut {NoStop}%
\bibitem [{\citenamefont {Li}\ and\ \citenamefont {Lee}(2018)}]{li2018a}%
  \BibitemOpen
  \bibfield  {author} {\bibinfo {author} {\bibfnamefont {X.}~\bibnamefont
  {Li}}\ and\ \bibinfo {author} {\bibfnamefont {S.}~\bibnamefont {Lee}},\
  }\href {\doibase 10.1103/PhysRevB.97.094309} {\bibfield  {journal} {\bibinfo
  {journal} {Phys. Rev. B}\ }\textbf {\bibinfo {volume} {97}},\ \bibinfo
  {pages} {094309} (\bibinfo {year} {2018})}\BibitemShut {NoStop}%
\bibitem [{\citenamefont {Yang}\ \emph {et~al.}(2019)\citenamefont {Yang},
  \citenamefont {Yue},\ and\ \citenamefont {Liao}}]{Nanalytical}%
  \BibitemOpen
  \bibfield  {author} {\bibinfo {author} {\bibfnamefont {R.}~\bibnamefont
  {Yang}}, \bibinfo {author} {\bibfnamefont {S.}~\bibnamefont {Yue}}, \ and\
  \bibinfo {author} {\bibfnamefont {B.}~\bibnamefont {Liao}},\ }\href {\doibase
  10.1080/15567265.2018.1551449} {\bibfield  {journal} {\bibinfo  {journal}
  {Nanosc. Microsc. Therm.}\ }\textbf {\bibinfo {volume} {23}},\ \bibinfo
  {pages} {25} (\bibinfo {year} {2019})}\BibitemShut {NoStop}%
\bibitem [{\citenamefont {Zhang}\ \emph {et~al.}(2019)\citenamefont {Zhang},
  \citenamefont {Guo},\ and\ \citenamefont {Chen}}]{ZHANG20191366}%
  \BibitemOpen
  \bibfield  {author} {\bibinfo {author} {\bibfnamefont {C.}~\bibnamefont
  {Zhang}}, \bibinfo {author} {\bibfnamefont {Z.}~\bibnamefont {Guo}}, \ and\
  \bibinfo {author} {\bibfnamefont {S.}~\bibnamefont {Chen}},\ }\href {\doibase
  10.1016/j.ijheatmasstransfer.2018.10.141} {\bibfield  {journal} {\bibinfo
  {journal} {Int. J. Heat Mass Transfer}\ }\textbf {\bibinfo {volume} {130}},\
  \bibinfo {pages} {1366} (\bibinfo {year} {2019})}\BibitemShut {NoStop}%
\bibitem [{\citenamefont {Barenblatt}(1987)}]{barenblatt1987dimensional}%
  \BibitemOpen
  \bibfield  {author} {\bibinfo {author} {\bibfnamefont {G.~I.}\ \bibnamefont
  {Barenblatt}},\ }\href@noop {} {\emph {\bibinfo {title} {Dimensional
  analysis}}}\ (\bibinfo  {publisher} {CRC Press},\ \bibinfo {year}
  {1987})\BibitemShut {NoStop}%
\bibitem [{\citenamefont {Majee}\ and\ \citenamefont
  {Aksamija}(2016)}]{PhysRevB.93.235423}%
  \BibitemOpen
  \bibfield  {author} {\bibinfo {author} {\bibfnamefont {A.~K.}\ \bibnamefont
  {Majee}}\ and\ \bibinfo {author} {\bibfnamefont {Z.}~\bibnamefont
  {Aksamija}},\ }\href {\doibase 10.1103/PhysRevB.93.235423} {\bibfield
  {journal} {\bibinfo  {journal} {Phys. Rev. B}\ }\textbf {\bibinfo {volume}
  {93}},\ \bibinfo {pages} {235423} (\bibinfo {year} {2016})}\BibitemShut
  {NoStop}%
\bibitem [{\citenamefont {Hale}\ and\ \citenamefont
  {Townsend}(2013)}]{NicholasH13GaussL}%
  \BibitemOpen
  \bibfield  {author} {\bibinfo {author} {\bibfnamefont {N.}~\bibnamefont
  {Hale}}\ and\ \bibinfo {author} {\bibfnamefont {A.}~\bibnamefont
  {Townsend}},\ }\href {\doibase 10.1137/120889873} {\bibfield  {journal}
  {\bibinfo  {journal} {Siam J. Sci. Comput}\ }\textbf {\bibinfo {volume}
  {35}},\ \bibinfo {pages} {A652} (\bibinfo {year} {2013})}\BibitemShut
  {NoStop}%
\end{thebibliography}%
\end{document}